\documentclass[a4paper,fleqn,usenatbib]{mnras}

\usepackage{newtxtext,newtxmath}
\usepackage{pdflscape}
\usepackage[T1]{fontenc}
\usepackage{ae,aecompl}
\usepackage{placeins}
\usepackage{enumitem}
\usepackage[dvipsnames]{xcolor}  % Added by AZ for having more colors

\usepackage{graphicx}	% Including figure files
\usepackage{amsmath}	% Advanced maths commands
\usepackage{amssymb}	% Extra maths symbols
\usepackage{soul}		% ADDED BY FB to cross out text
\setstcolor{red}
\setul{}{0.5mm}		        % \setul{⟨underline depth⟩}{⟨underline thickness⟩}
\def\hi{\relax \ifmmode {\mbox H\,{\scshape i}}\else H\,{\scshape i}\fi}
\def\hii{\relax \ifmmode {\mbox H\,{\scshape ii}}\else H\,{\scshape ii}\fi}
\def\hei{\relax \ifmmode {\mbox He\,{\scshape i}}\else He\,{\scshape i}\fi}
\def\heii{\relax \ifmmode {\mbox He\,{\scshape ii}}\else He\,{\scshape ii}\fi}
\def\nii{\relax \ifmmode {\mbox N\,{\scshape ii}}\else N\,{\scshape ii}\fi}
\def\ni{\relax \ifmmode {\mbox N\,{\scshape i}}\else N\,{\scshape i}\fi}
\def\oi{\relax \ifmmode {\mbox O\,{\scshape i}}\else O\,{\scshape i}\fi}
\def\oii{\relax \ifmmode {\mbox O\,{\scshape ii}}\else O\,{\scshape ii}\fi}
\def\oiii{\relax \ifmmode {\mbox O\,{\scshape iii}}\else O\,{\scshape iii}\fi}
\def\sii{\relax \ifmmode {\mbox S\,{\scshape ii}}\else S\,{\scshape ii}\fi}
\def\siii{\relax \ifmmode {\mbox S\,{\scshape iii}}\else S\,{\scshape iii}\fi}
\def\ariii{\relax \ifmmode {\mbox Ar\,{\scshape iii}}\else Ar\,{\scshape iii}\fi}
\def\ariv{\relax \ifmmode {\mbox Ar\,{\scshape iv}}\else Ar\,{\scshape iv}\fi}
\def\neiii{\relax \ifmmode {\mbox Ne\,{\scshape iii}}\else Ne\,{\scshape iii}\fi}
\def\cliii{\relax \ifmmode {\mbox Cl\,{\scshape iii}}\else Cl\,{\scshape iii}\fi}
\def\feiii{\relax \ifmmode {\mbox Fe\,{\scshape iii}}\else Fe\,{\scshape iii}\fi}
\def\feii{\relax \ifmmode {\mbox Fe\,{\scshape ii}}\else Fe\,{\scshape ii}\fi}
\def\niqii{\relax \ifmmode {\mbox Ni\,{\scshape ii}}\else Ni\,{\scshape ii}\fi}
\def\cii{\relax \ifmmode {\mbox C\,{\scshape ii}}\else C\,{\scshape ii}\fi}
\def\mgi{\relax \ifmmode {\mbox Mg\,{\scshape i}}\else Mg\,{\scshape i}\fi}
\def\silii{\relax \ifmmode {\mbox Si\,{\scshape ii}}\else Si\,{\scshape ii}\fi}
\def\ha{\relax \ifmmode {\mbox H}\alpha\else H$\alpha$\fi}
\def\hb{\relax \ifmmode {\mbox H}\beta\else H$\beta$\fi}
\def\me{$^{-1}$}
\def\arcsec{\hbox{$^{\prime\prime}$}}

\def\deg{\hbox{$^{\circ}$}}
\defcitealias{paperI}{Paper~I}
\title[Bar effect on abundance gradients. II. Luminosity-dependent flattening]{Bar effect on gas-phase abundance gradients. II. Luminosity-dependent flattening}
\author[A. Zurita et al.]{A. Zurita,$^{1,2}$\thanks{E-mail: azurita@ugr.es}
E. Florido,$^{1,2}$
F. Bresolin,$^{3}$
I. P\'erez$^{1,2}$
and E. P\'erez-Montero$^{4}$
\\
$^{1}$Dpto. de F\'\i sica y del Cosmos, Campus de Fuentenueva, Edificio Mecenas, Universidad de Granada, E18071--Granada, Spain\\
$^{2}$Instituto Carlos I de F\'\i sica Te\'orica y Computacional, Facultad de Ciencias, E18071--Granada, Spain\\
$^{3}$Institute for Astronomy, 2680 Woodlawn Drive, Honolulu, HI 96822, USA\\
$^{4}$Instituto de Astrof\' isica de Andaluc\' ia, Camino Bajo de Hu\'etor s/n, Aptdo. 3004, E18080-Granada, Spain\\
}

\date{Accepted XXX. Received YYY; in original form ZZZ}

\pubyear{2020}

\usepackage{etoolbox}
\makeatletter
\patchcmd\@combinedblfloats{\box\@outputbox}{\unvbox\@outputbox}{}{\errmessage{\noexpand patch failed}}
\makeatother

\begin{document}
\label{firstpage}
\pagerange{\pageref{firstpage}--\pageref{lastpage}}
\maketitle

% Abstract of the paper
\begin{abstract}
%Stellar bars are believed to be major drivers of the secular evolution of disc galaxies. The bar-induced gas flows within discs  mix the gas and can flatten radial gas-phase abundance profiles according to theoretical work, but observational studies  report contradicting results. 
We present here the second part of a project that aims at solving the controversy on the issue of the bar effect on the radial distribution of metals in the gas-phase of spiral galaxies. In Paper I we presented a compilation of more than 2800 \hii\ regions belonging to 51 nearby galaxies for which we derived chemical abundances and radial abundance profiles from a homogeneous methodology. In this paper we analyse the derived gas-phase radial abundance profiles of 12+$\log$(O/H) and $\log$(N/O), for barred and unbarred galaxies separately, and find that the  differences in slope between barred and unbarred galaxies depend on  galaxy luminosity. This is due to a different dependence of the abundance gradients (in dex~kpc\me) on luminosity for the two types of galaxies: In the galaxy sample that we consider the gradients appear to be considerably shallower for strongly barred galaxies in the whole luminosity range, while profile slopes for unbarred galaxies become steeper with decreasing luminosity. Therefore, we  only detect differences in slope for the lower luminosity (lower mass) galaxies  (M$_B \gtrsim -19.5$ or $M_*\lesssim10^{10.4}$~M$_\odot$).  We discuss the results in terms of the disc evolution and radial mixing induced by bars and spiral arms. Our results reconcile previous discrepant findings that were biased by  the luminosity (mass) distribution of the sample galaxies and possibly by the abundance diagnostics employed.
%The results presented here provide key elements for understanding the chemical evolution of galaxies.
\end{abstract}

% Select between one and six entries from the list of approved keywords.
% Don't make up new ones.

%%%%%%%%%%%%%%%%%%%%%%%%%%%%%%%%%%%%%%%%%%%%%%%%%%
%%%%%%%%%%%%%%%%% BODY OF PAPER %%%%%%%%%%%%%%%%%%
%%%%%%%%%%%%%%%%%%%%%%%%%%%%%%%%%%%%%%%%%%%%%%%%%%

\section{Introduction}
One of the major challenges in Astrophysics is understanding galaxy evolution and, in particular, the relative importance of external and internal processes on such evolution and therefore on the present properties of galaxies. In the current paradigm, the evolution of galaxies was rapid and violent at early times, driven by external processes such as mergers and galaxy interactions. Later on, additional and slower processes, the so-called {\em secular} processes, came into play and will become dominant in the future \cite[e.g.][]{kormendy}.

Galactic bars, or simply {\em bars}, are considered key agents for the internally-driven secular evolution of disc galaxies.  Bars are especially prominent in optical and infrared images as elongated luminous structures in the central regions of $\sim$30-70\% of disc galaxies \citep[e.g.][]{nair,Masters2011,Menendez-Delmestre07}, with semi-major axes of up to ten kiloparsecs \citep{erwin}.
%, and are frequent among disk galaxies, with and observed fraction of $\sim$30-70\% depending on redshift and observational photometric band \citep[e.g.][]{nair,Masters2011,Menendez-Delmestre07}.%Sheth+2008 
  The non-axisymmetric distribution of matter of the bar produces a gravitational potential that redistributes stars and gas inside the disc by pushing them inwards (outwards) within (beyond) the disc corotation radius \citep[e.g.][]{SW93,athanassoula13}. This bar-induced matter redistribution might secularly drive evolution processes in galaxies.

%Such matter redistribution might secularly affect the galaxy evolution. %% Citas a papers de modelos

Observationally, it has been difficult to find strong evidence for the bar-induced secular evolution of galaxies, in part due to the difficulty in isolating the effects of a bar from others that also contribute to  matter rearrangement inside galaxies (e.g.  minor mergers or tidal interactions). In spite  of these difficulties there is increasing observational evidence that bars produce the build-up of pseudo-bulges \citep[e.g.][]{cheung,Kruk2018}. In addition, in galaxy centres bars increase the concentration of molecular gas \citep[e.g.][]{sakamoto,sheth,jogee2005} and enhance the star formation rate  \citep[e.g.][]{Ho_barras,ellison,ooy,wang2012,bulbos,Chown} and, within the discs, possibly induce the cessation of the  star formation \citep[by the so called {\em bar quenching}, e.g.][]{cheung,Kruk2018,Newnham2020}.

One important consequence of the bar-induced large-scale gas flows in disc galaxies \citep[e.g.][]{SW93,athanassoula03} is the mixing of the gas with presumably different chemical abundances. Therefore, according to simulations bars can flatten the observed negative radial profile of  12+$\log$(O/H) (or {\em metallicity} profile) in spirals \citep[e.g.][]{pilyugin2014,Sanchez-Menguiano2016,Bresolin2019,Perez-Montero2016}, as  inwards gas flows dilute the higher metal content in the central regions, while outwards flows can enrich the more metal-poor outer disc areas.

However, this theoretically predicted impact of bars on the gas-phase radial metallicity profiles has been  difficult to confirm through observations, and contradicting results can be easily found in the literature. \citet{pagel79} first suggested that the  bar could be responsible for the shallow radial metallicity profile observed in NGC~1365, and this work inspired many other studies in the 1990s. These tried to investigate the relationship between gas metallicity gradients in spirals and macroscopic properties of their host galaxies, among them the presence of a stellar bar. \citet{vila-costas92} and \citet{Zaritsky94} reported a trend for barred galaxies to have shallower metallicity gradients than unbarred galaxies from long-slit spectroscopy of \hii\ regions in nearby galaxies, but the number of barred galaxies ($\sim6$\,-\,$7$) was small in their data samples. The samples were enlarged to comprise up to  $\sim16$ barred galaxies in the work developed by \citet{Martin94} and \citet{Dutil_Roy}, who corroborated the flattening effect of bars on radial metallicity gradients from additional spectrophotometry and a compilation of slopes derived by previous authors. These authors also claimed the existence of a relation between gas-phase metallicity gradients and  bar properties, with stronger bars associated to shallower gradients.
%performed a morphological characterization of the bars and claimed a relation among the metallicity gradients and the bar properties, with stronger bars showing shallower gradients. %Roy97,Dutil_Roy

This research topic remained rather inactive until the advent of the Integral Field Unit (IFU) spectroscopy instruments and their associated surveys, that considerably improved the number statistics and the spatial coverage of galactic discs. IFU-based works, such as those exploiting CALIFA data \citep{Walcher2014}, with sub-samples of 66 to 201 galaxies, compared gas-phase metallicity gradients of barred and unbarred galaxies and found no significant difference \citep{sanchez14,Sanchez-Menguiano2016, Zinchenko2019,Perez-Montero2016}, contradicting the results from the 1990s. In addition to  O/H, \cite{Perez-Montero2016} also analysed the N/O radial profile slopes and found no significant difference either. The VENGA IFU survey \citep{blanc2013} has also been used to investigate this issue with improved spatial resolution with respect to CALIFA data, but with a much smaller sample of only eight galaxies, reporting also no difference in O/H radial profiles between  barred and unbarred galaxies \citep{Kaplan2016}.

The controversy on the effects of bars on chemical abundances is not limited to the metallicity gradients, but also involves the 
values in the central regions.  The gas-phase central metallicity of barred and unbarred galaxies was analysed by \cite{ellison} and \cite{cacho} from SDSS spectra, with the former finding larger central (inner few kpc) oxygen abundances in barred galaxies with respect to unbarred ones, while the latter found no difference. A subsequent work by \citet{bulbos}, also based on SDSS spectra, found no difference in central O/H between barred and unbarred galaxies, but a significant difference in central nitrogen-to-oxygen abundance ratios, with barred galaxies showing an enhanced N/O ratio with respect to unbarred galaxies, notably in the lower mass range ($\lesssim10^{10.4}$M$_\odot$). The discrepant results between  this study and the works by \cite{ellison} and \cite{cacho} were explained by \citet{bulbos} in terms of sample selection issues and choice of calibrations of strong-line nebular abundance methods based on [\nii] emission lines \citep[i.e.][]{pmc09}.

There are important differences between the previous (discrepant) works, namely the methodology in the derivation of the chemical abundances,  the number statistics of the data samples and the spatial resolution of the data. In fact, these different investigations were based on either a  {\em spaxel-by-spaxel}  analysis, spatially resolved spectroscopy of individual \hii\ regions, or the integrated spectral study of ensembles of star-forming sites. This heterogeneity, together with considerations concerning the important role that bars may play on disc galaxy evolution,  motivated us to revisit the  issue of the bar effects on the gas-phase metallicity distribution in disc galaxies. The work presented here is the second part of the project, for which we did a major compilation of emission-line fluxes of \hii\ regions in a  sample of barred and unbarred spirals. In \citet{paperI}, hereinafter  referred to as \citetalias{paperI}, we presented the \hii\ region and galaxy samples and obtained chemical abundances and radial O/H and N/O abundance profiles using a homogeneous methodology. This paper concentrates on the analysis of the radial abundance profiles derived for barred and unbarred galaxies separately, and it is organized as follows. Sect.~\ref{data} summarises the main details of  the galaxy and \hii\ region samples. In Sect.~\ref{intercepts} we analyse the gas-phase radial abundance profiles comparatively for (strongly and weakly) barred and unbarred galaxies. The metallicity gradients and central abundances of barred and unbarred galaxies are also analysed as a function of luminosity and morphological type. The dependence of slopes and intercepts of linear regressions to the $\log$(O/H)  and $\log$(N/O) profiles with bar parameters is explored in Sects.~\ref{bar_params} and \ref{additional_considerations}. Sect.~\ref{discusion} contains our discussion and  we present a summary and our concluding remarks in Sect.~\ref{conclusiones}.

%%%%%%%%%%%%%%%%%%%%%%%%%%%%%%%%%%%%%%%%%%%%%%%%%%%%%%%%%%%%%%%%%%%%%%%%%%%%%%%%%%%%%%%%%%%%%%%%%%%%%%%%%%%%%%
%%%%%%%%%%%%%%%%%%%%%%%%%%%%%%%%%%%%%%%%%%%%%%%%%%%%%%%%%%%%%%%%%%%%%%%%%%%%%%%%%%%%%%%%%%%%%%%%%%%%%%%%%%%%%%
%%%%%%%%%%%%%%%%%%%%%%%%%%%%%%%%%%%%%%%%%%%%%%%%%%%%%%%%%%%%%%%%%%%%%%%%%%%%%%%%%%%%%%%%%%%%%%%%%%%%%%%%%%%%%%
\section{Data and methodology}
\label{data}
The data and the methodology employed for the derivation of the galactic structural parameters, nebular O/H and N/O abundance ratios,  the radial abundance profiles and the corresponding linear fits were presented in \citetalias{paperI}. We summarise here the basic information and refer the reader to this paper for further details.

\subsection{Galaxy sample and bar classes}
\label{galaxy_sample}
The galaxy sample was presented in \citetalias{paperI} and comprises 51 nearby spirals (distances$<64$~Mpc), with inclination angles $i<70$\deg, for which emission-line ratios from resolved spectroscopy of \hii\ regions and celestial coordinates were available from previous publications. We imposed the availability of these data for at least seven \hii\ regions covering a wide range in galactocentric distance in order to derive chemical abundance radial profiles reliably.  The $B$-band absolute magnitude of the galaxies (M$_B$) ranges from $-17$ to $-22$. Our final sample comprises 22 strongly barred, 9 weakly barred and 20 unbarred galaxies, i.e. there is virtually equal representation of strongly barred and unbarred systems. Also, barred and unbarred galaxies cover a similar parameter space and have similar distributions in terms of M$_B$, disc inclination, morphological T-type and disc effective radius \citepalias[see fig.~2 in][]{paperI}.

The galaxies were classified into bar classes according to their deprojected bar ellipticity, \citep[$e_{bar,d}$, see e.g.][]{abraham00}, as this parameter correlates with the bar gravitational torque \citep{SimonDiaz2016}. Galaxies were considered to be strongly barred when $e_{bar,d} \geqslant 0.5$ and weakly barred when $0.3\leqslant e_{bar,d} < 0.5$.  If a galaxy has a detected central oval distortion with deprojected ellipticity smaller than 0.3, then it is considered as unbarred.

The bar ellipticity ($e_{bar}$), together with other bar parameters (such as bar length and position angle) and disc parameters -- disc scale length (r$_d$),  disc effective radius\footnote{For an exponential disc profile, r$_e$ =1.678~r$_d$, with r$_d$ the disc scale length.} ($r_e$), PA and inclination -- were derived from compiled broad-band images\footnote{Mostly Sloan $r$-band or Johnson R-band images, except for two galaxies. See \citetalias{paperI} for further details.} of all galaxies in the sample (except for M31 and the Milky Way). We used  the method of fitting ellipses to the image isophotes. We refer the reader to \citetalias{paperI} for detailed information on the galaxy sample and on the methodology adopted for the determination of these parameters, together with a comparison between our bar classification and that obtained from the RC3 catalogue \citep{RC3}.

\subsection{H\,{\small II} region sample and chemical abundances}
\label{abundances}
We performed a compilation of emission-line fluxes and celestial coordinates for observed \hii\ regions of the sample galaxies described in Sect.~\ref{galaxy_sample}. The compilation comprises 2831 independent measurements and includes fluxes, normalised to those of the H$\beta$ line, for the brightest emission lines ([\oii]$\, \lambda\lambda$3726,3729,  [\oiii]$\,\lambda\lambda$4959,5007, [\nii]$\,\lambda\lambda$6548,6583, \ha, [\sii]$\,\lambda\lambda$6717,6731, [\siii]$\,\lambda\lambda$9069,9532). The auroral-line fluxes ([\sii]$\,\lambda\lambda$4068,4076,  [\oiii]$\,\lambda$4363, [\nii]$\lambda$5755, [\siii]$\,\lambda$6312 and [\oii]$\,\lambda\lambda$7320,7330) were also compiled when available (for 709 out of the 2831). From the 
compiled extinction-corrected fluxes we derived 12$+\log$(O/H) and $\log$(N/O) from a variety of methods: (a) The $T_e$-based or {\em direct} method was employed for the subsample of regions with auroral-line fluxes, which allowed us to determine  {\em direct} abundances for 610 \hii\ regions. (b) For the whole \hii\ region sample we used a variety of strong-line methods for the derivation of 12$+\log$(O/H), such as HII-CHI-mistry method (hereinafter HCM) \citep[][hereinafter PM14]{epm14}, O3N2 as calibrated by \citet[][hereinafter PP04]{pp04}, and later by \citet[][hereinafter  M13]{marino13}, N2 with the empirical calibration given by \citet{pp04},  N2O2  with the  empirical calibration by \citet[][hereinafter B07]{b07}, the $R$ calibration by \citet[][hereinafter PG16]{PilyuginGrebel2016}, and R$_{23}$, as calibrated from  theoretical model grids by \citet[][hereinafter MG91]{m91}, with the \citet{kkp99} parametrization. The  N/O abundance ratio was derived with HCM, but we also used the empirical calibrations of N2O2 and N2S2  provided by \citet[PMC09,][]{pmc09} and the $R$ calibration \citep{PilyuginGrebel2016}.

\subsection{Radial abundance profiles}
\label{fits}
We derived the radial 12$+\log$(O/H) and $\log$(N/O) profiles from our recalculated abundances for the \hii\ regions and from their deprojected galactocentric distances (utilizing the location information provided by the different authors together with our derived disc structural parameters for the host galaxies). For about 75\% of the galaxies the profiles extend to the isophotal radius r$_{25}$ or beyond, while in six galaxies (12\% of the sample) the collected data extend out only to $\sim$0.6-0.8 times r$_{25}$.  The 12+$\log$(O/H) and $\log$(N/O) \hii\ region abundances  as a function of galactocentric radius ($R$) were presented in \citetalias{paperI} for all galaxies, and were fitted with a least-squares linear regression as parametrized by 12+$\log$(O/H)$=b_{O/H} + \alpha_{O/H} R$ and $\log$(N/O)=$b_{N/O} + \alpha_{N/O} R$, respectively. Circumnuclear \hii\ regions, regions showing signs of contamination from  hard ionization sources or shock-excitation, and regions located in the central areas of galaxies hosting an AGN were not used for the fits, together with \hii\ regions showing a systematically larger deviation from the bulk of the data \citepalias[see sect.~5 in][for further details]{paperI}.

The radial abundance profiles for both the strong-line and {\em direct}  abundances were then characterised by their corresponding y-intercepts and slopes ($\alpha_{N/O}$ and $\alpha_{O/H}$), where the latter were also calculated for radial profiles normalised to $r_e$ and $r_{25}$, in order to  facilitate comparisons with results from other authors. We used the reduced chi-squared, $\chi_\nu^2$, statistics in order to judge the significance of radial breaks in the abundance profiles: when the double linear fit implies an improvement in $\chi_\nu^2$ of at least 10\%, separate inner and outer slopes were also calculated. In what follows, unless specified, we will always refer to the gradients or slopes corresponding to the single linear fits to the radial abundance profiles.

In \citetalias{paperI} we performed a detailed evaluation of the strong-line-based methods for abundance determinations in relation to the $T_e$-based one,  both for  O/H and N/O abundances for individual  \hii\ regions, and for the abundance gradients derived from both methodologies. As we aim at solving the controversy emerged from previous works, we analyse here slopes and y-intercepts for all the strong-line abundance determination methods, even though some of them have been shown to depart from the $T_e$-based  scale.
%% Except for N/O, for which we will only show results as obtained with the HCM method for the reasons explained in \citetalias{paperI}.

%%%%%%%%%%%%%%%%%%%%%%%%%%%%%%%%%%%%%%%%%%%%%%%%%%%%%%%%%%%%%%%%%%%%%%%%%%%%%%%%%%%%%%%%%%%%%%%%%%%%%%%%%%%%%%
%%%%%%%%%%%%%%%%%%%%%%%%%%%%%%%%%%%%%%%%%%%%%%%%%%%%%%%%%%%%%%%%%%%%%%%%%%%%%%%%%%%%%%%%%%%%%%%%%%%%%%%%%%%%%%
%%%%%%%%%%%%%%%%%%%%%%%%%%%%%%%%%%%%%%%%%%%%%%%%%%%%%%%%%%%%%%%%%%%%%%%%%%%%%%%%%%%%%%%%%%%%%%%%%%%%%%%%%%%%%%

\section{Intercepts and slopes for barred/unbarred galaxies}
\label{intercepts}

\begin{figure*}
\begin{minipage}{1.0\textwidth}
\hspace{-0.8cm}\includegraphics[width=0.58\textwidth, clip, trim = 0cm 0.0cm 0cm 0.0cm]{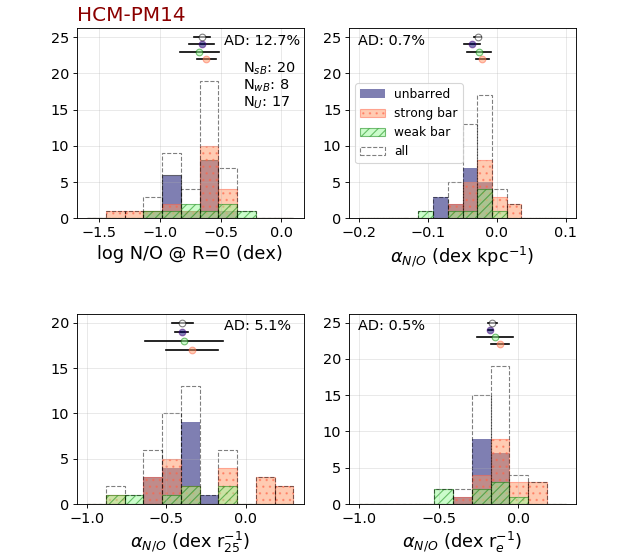} 
\hspace{-0.8cm}\includegraphics[width=0.58\textwidth, clip, trim = 0cm 0.0cm 0cm 0.0cm]{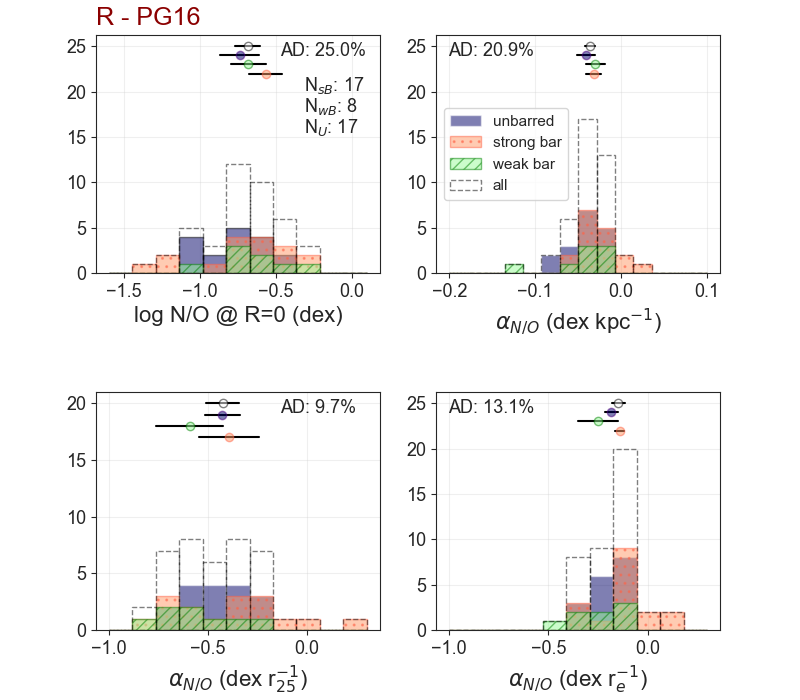}
\vspace{-0.3cm}
\caption{Histograms showing y-intercepts  ({\em top-left}) and slopes of the linear fits to the radial $\log$(N/O) abundance profiles expressed in dex~kpc$^{-1}$ ({\em top-right}), in units of r$_{25}$ ({\em bottom-left}) and in units of the disc effective radius ({\em bottom-right}), for strongly barred (dotted orange), weakly barred (dashed green) and unbarred (blue) galaxies separately, for abundances obtained from HCM (left) and the $R$ calibration (right). The number of galaxies in each sub-sample and the $P$-value for the two-sample Anderson-Darling test (AD in \%) for the distribution of strongly barred and unbarred galaxies are also shown. The dashed black line shows the histogram for all galaxies in the sample. The circles in the upper side of the panels indicate the median value for each distribution and the horizontal error bar covers the 95\% confidence interval for the corresponding median value. \label{histog_fits_HCM_NO}}
\end{minipage}
\end{figure*}

\begin{figure*}
\begin{minipage}{1.0\textwidth}
\hspace{-0.8cm}\includegraphics[width=0.58\textwidth]{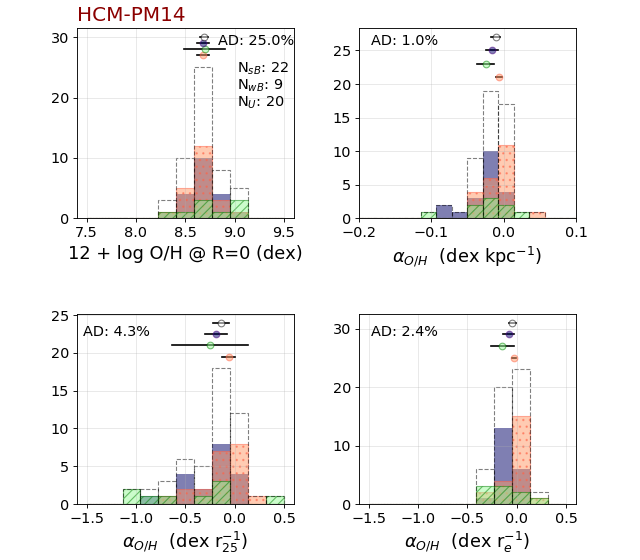}
\hspace{-0.8cm}\includegraphics[width=0.58\textwidth]{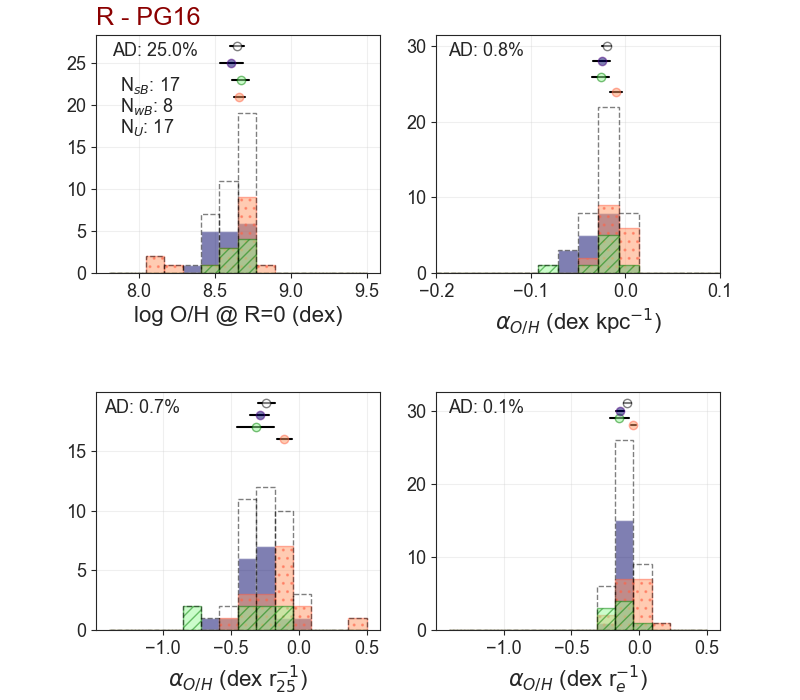}
\end{minipage}
\caption{Same as Fig.~\ref{histog_fits_HCM_NO} but for 12+$\log$(O/H) as obtained from HCM (left) and the $R$ calibration (right). \label{histog_fits_HCM_OH}}
\end{figure*}

\begin{figure*}
\begin{minipage}{1.0\textwidth}
\hspace{-0.8cm}\includegraphics[width=0.58\textwidth]{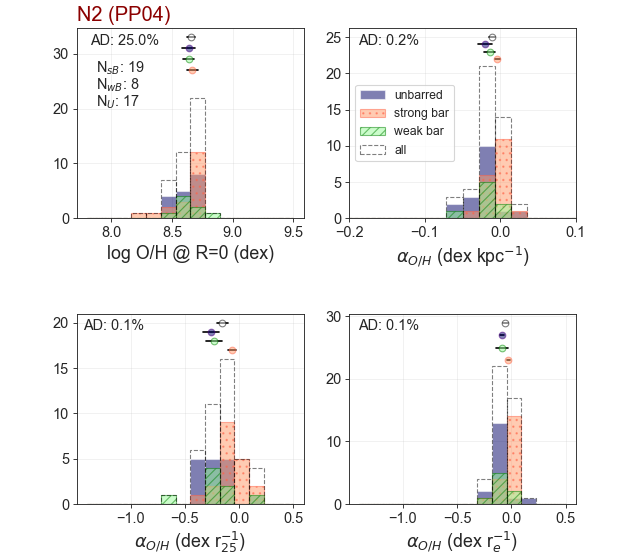}
\hspace{-0.8cm}\includegraphics[width=0.58\textwidth]{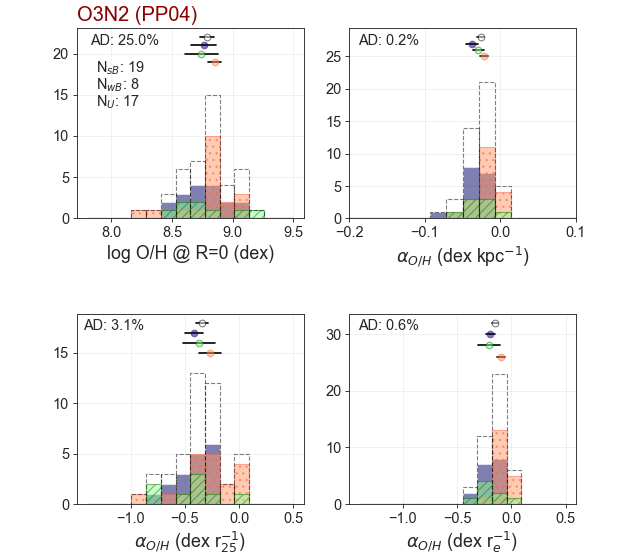}

\vspace{0.8cm}

\hspace{-0.8cm}\includegraphics[width=0.58\textwidth]{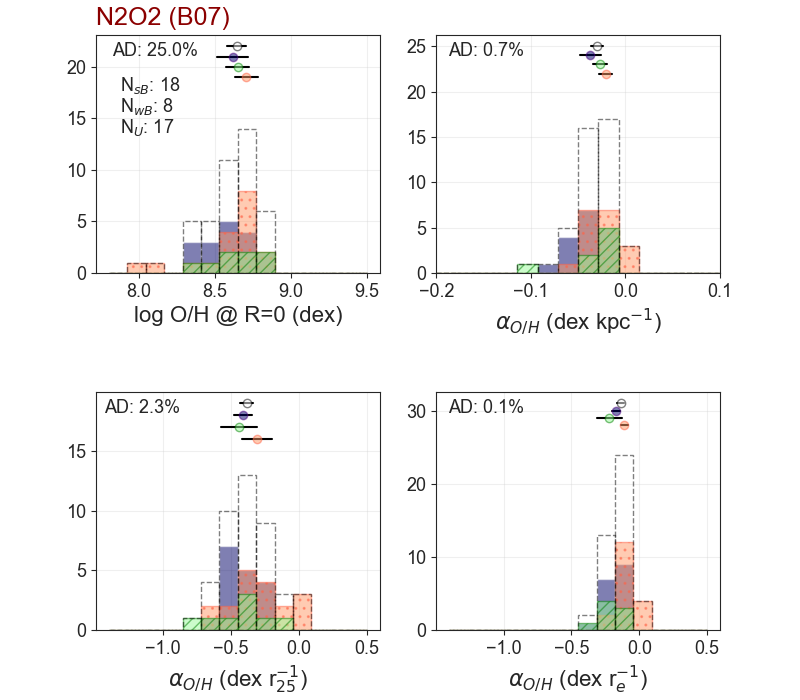}
\hspace{-0.8cm}\includegraphics[width=0.58\textwidth]{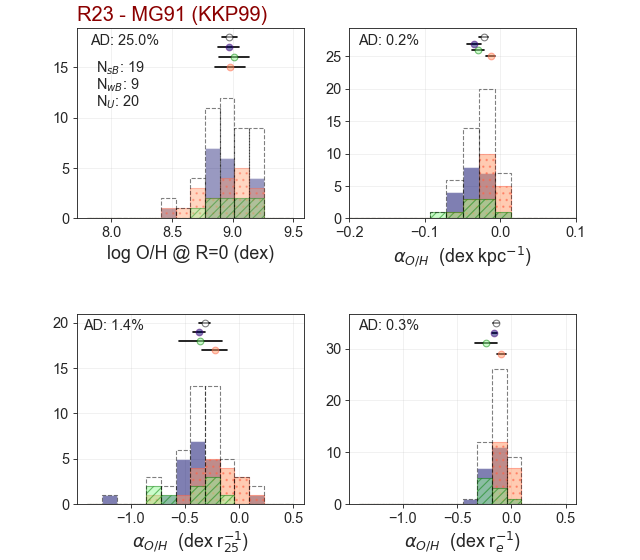}
\caption{Histograms showing central 12+$\log$(O/H) abundances  and radial abundance gradients with different normalizations for some of the strong-line methods used in this work. From top to bottom and from left to right: N2 (PP04), O3N2 (PP04), N2O2 (B07), R$_{23}$ (MG91). Colours, symbols and information in plots as in Figs.~\ref{histog_fits_HCM_NO} and \ref{histog_fits_HCM_OH}. \label{histog_empi}}
\end{minipage}
\end{figure*}

Figs.~\ref{histog_fits_HCM_NO} to \ref{histog_empi} show histograms of  the y-intercepts  and the slopes of the $\log$(N/O) and  $12+\log$(O/H) radial abundance profiles (from strong-line methods) for different normalizations and abundance strong-line methods, as summarised in Sect.~\ref{abundances}.
 
The central $\log$(N/O) abundances, as traced by the y-intercepts of the radial fits, $b_{N/O}$, range from  $\sim -1.3$~dex for NGC~1313 and NGC~4395 
to $\sim -0.25$~dex, considering  the complete sample of galaxies for which the N/O radial profile could be derived from HCM and the $R$ calibration. These are the methods that better reproduce the $T_e$-based scale for our galaxy sample (see \citetalias{paperI}). Table~\ref{tab:fits_NO} shows the median values of the central $\log$(N/O) of the different sub-samples, indicated with circles in the histograms in the top-left panels for HCM and the R calibration in  Fig.~\ref{histog_fits_HCM_NO}. The central $\log$(N/O) is marginally larger for strongly barred than for unbarred galaxies for all the strong-line methods employed for N/O, but differences are within the error bars. The two-sample Anderson-Darling (hereinafter AD) test output yields, for all the methods (although N2O2 and N2S2 are not shown in the histograms), a $P$-value above the 5\% threshold normally adopted, implying that the distributions for strongly barred and unbarred galaxies come from the same parent distributions. The same result is obtained when comparing all (strongly and weakly) barred galaxies with the unbarred ones.

No difference is observed either in the central 12+$\log$(O/H) abundance of strongly  barred and unbarred galaxies, regardless of the strong-line method employed, as shown in Table~\ref{tab:fits_OH} and in the top-left panels of Figs.~\ref{histog_fits_HCM_OH} and \ref{histog_empi}. In addition, the AD test $P$-value is in all cases considerably above 5\% implying similar distributions in central values for strongly barred and unbarred galaxies.

\begin{table*}
\centering
\caption{Median value (and corresponding error) of the N/O central abundance ratio  (y-intercept from the radial profile fitting) and slope of the radial abundance gradient in different radial units for (strongly and weakly) barred and unbarred galaxies separately and for all the galaxies of the sample. N indicates the number of galaxies in each subsample.  We include the results obtained using all the different strong-line techniques adopted for N/O.}
\label{tab:fits_NO}
\begin{tabular}{lccccc} 
 \hline
	                               &  log(N/O)            &  $\alpha_{N/O}$             &  $\alpha_{N/O}$       & $\alpha_{N/O}$     & N   \\ 
                                       &  at  R=0              &  (dex kpc$^ {-1}$)          &  (dex r$_{25}^ {-1}$)  & (dex r$_{e}^ {-1}$) &   \\ 
\hline
\multicolumn{6}{l}{HCM - \citet{epm14}}\\
\hline
strongly barred               & $-0.62\pm0.08$  &   $-0.021\pm0.009$   &  $-0.3\pm0.2 $     & $ -0.11\pm0.06$ & 20 \\ %% Revisado el 06/07/20
weakly   barred               & $-0.7\pm0.2 $     &   $-0.03\pm0.02$       &  $-0.4\pm0.2 $     & $ -0.1\pm0.1$     &  8 \\
strong + weakly barred  & $-0.65\pm0.08$  &   $-0.022\pm0.007$   & $ -0.3\pm0.1 $     & $ -0.11\pm0.04$ & 28 \\
unbarred                         & $-0.7\pm0.1$      &  $ -0.04\pm0.01 $      &  $-0.40\pm0.04$   & $ -0.18\pm0.02$ & 17 \\
all                                   & $-0.65\pm0.07$  &   $-0.028\pm0.005$   &  $-0.40\pm0.07 $  & $ -0.16\pm0.03$ & 45 \\
\hline
\multicolumn{6}{l}{$R$ calibration - \citet{PilyuginGrebel2016}}\\
\hline 
strongly barred               & $-0.6\pm0.1$     &   $-0.032\pm0.009$   &  $-0.4\pm0.2 $      & $ -0.14\pm0.02$ & 17 \\ %% Revisado el 06/07/20
weakly   barred               & $-0.7\pm0.1 $     &   $-0.03\pm0.01$       &  $-0.6\pm0.2 $      & $ -0.3\pm0.1$     &  8 \\
strong + weakly barred  & $-0.67\pm0.09$  &   $-0.030\pm0.007$   & $ -0.4\pm0.1 $      & $ -0.14\pm0.05$ & 25 \\
unbarred                         & $-0.7\pm0.1$      &  $ -0.04\pm0.01 $      &  $-0.43\pm0.09$   & $ -0.18\pm0.03$ & 17 \\
all                                   & $-0.69\pm0.08$  &   $-0.036\pm0.006$   &  $-0.43\pm0.08 $  & $ -0.15\pm0.04$ & 42 \\
\hline
\multicolumn{6}{l}{N2O2 - \citet{pmc09}}\\
\hline 
strongly barred               & $-0.2\pm0.1$      &   $-0.04\pm0.01$      &  $-0.6\pm0.2 $       & $ -0.18\pm0.06$ & 18 \\ %% Revisado el 06/07/20
weakly   barred               & $-0.3\pm0.1 $     &   $-0.044\pm0.009$  &  $-0.7\pm0.2 $       & $ -0.4\pm0.2$     &  8 \\
strong + weakly barred  & $-0.3\pm0.1$      &   $-0.04\pm0.01$      & $ -0.4\pm0.1 $       & $ -0.20\pm0.07$ & 26 \\
unbarred                         & $-0.4\pm0.2$      &  $ -0.07\pm0.03 $    &  $-0.71\pm0.09$    & $ -0.28\pm0.04$ & 17 \\
all                                  & $-0.3\pm0.1$      &   $-0.05\pm0.01$      &  $-0.64\pm0.09 $   & $ -0.25\pm0.05$ & 43 \\
\hline
\multicolumn{6}{l}{N2S2 - \citet{pmc09}}\\
\hline 
strongly barred                & $-0.5\pm0.1$       &   $-0.02\pm0.01$        &  $-0.3\pm0.2 $      & $ -0.09\pm0.05$ & 20 \\ %% Revisado el 06/07/20
weakly   barred                & $-0.4\pm0.2 $      &   $-0.03\pm0.02$        &  $-0.4\pm0.2 $      & $ -0.2\pm0.1$     &  8 \\
strong + weakly barred   & $-0.5\pm0.1$       &   $-0.03\pm0.01$        & $ -0.4\pm0.1 $      & $ -0.10\pm0.05$ & 28 \\
unbarred                         & $-0.56\pm0.07$    &  $ -0.03\pm0.02 $      &  $-0.40\pm0.09$   & $ -0.17\pm0.04$ & 17 \\
all                                   & $-0.52\pm0.06$    &   $-0.026\pm0.006$   &  $-0.38\pm0.09 $  & $ -0.14\pm0.04$ & 45  \\
\hline 
\end{tabular}
\end{table*}
The comparison of the radial N/O and O/H abundance gradient slopes between strongly barred and unbarred galaxies does show clear differences between the two  types of galaxies. This can be seen in Fig.~\ref{histog_fits_HCM_NO} and Table~\ref{tab:fits_NO} for N/O, and in Figs.~\ref{histog_fits_HCM_OH} and \ref{histog_empi} and Table~\ref{tab:fits_OH} for all  the estimation methods we used. For both N/O and O/H, strongly barred galaxies show a shallower radial gradient ($\alpha_{N/O}$ and $\alpha_{O/H}$  closer to zero) than unbarred galaxies. This trend is observed for all normalisations of the slope, with AD $P$-values between strongly barred and unbarred galaxies well below 5\%, in fact below 1\% in most cases, except for N/O as derived from the $R$ calibration.

Slopes normalised to the isophotal radius r$_{25}$ show generally a larger dispersion, and the distributions of slopes in dex~r$_{25}^{-1}$  for strongly barred and unbarred galaxies yield slightly larger AD $P$-values. These $P$-values still indicate statistically significant differences between strongly barred and unbarred galaxies, except in the case of the O/H abundance gradients obtained with O3N2 as calibrated by M13 (not shown in the figures).
%obtained with  HCM and O3N2 as calibrated by M13 (the latter not shown in the figures). %MOdifico texto tras corregir error en asignacion de tipo de barra de MW y N1068. Tras repetir graficas, con HCM AD menor del 5%.

We have also done the exercise of comparing the distributions of radial abundance slopes for the sample of barred galaxies, including those that are either  weakly or  strongly barred, with the sample of unbarred galaxies. The trend for barred galaxies to show shallower slopes than unbarred galaxies is maintained when we include weakly barred galaxies. However, the differences in median values are considerably reduced, and remain within the errors. The AD test also yields larger $P$-values but still below 5\% in most  cases when the slopes are expressed in dex~kpc\me\ or dex~r$_{e}^{-1}$.  The result is reasonable. On the one hand, the sub-sample of weakly barred galaxies is smaller ($\sim1/3$ of the strongly barred galaxies) and, on the other hand, the dispersion of their slopes is considerably larger than for the unbarred or strongly barred galaxies alone. Therefore the inclusion of weakly barred galaxies in the comparison smooths out the differences.

In summary, strongly barred galaxies seem to have shallower radial abundance gradients than unbarred galaxies for both  $\log$(N/O) and 12+$\log$(O/H). The differences are more evident when slopes are expressed in  dex~kpc$^{-1}$ or  dex~$r_e^{-1}$, as the dispersion is considerably larger for slopes expressed in  dex~$r_{25}^{-1}$. No difference is observed in the central 12+$\log$(O/H) and $\log$(N/O) values, as given by the y-intercepts.
%We also observe a trend for strongly barred galaxies to have larger central $\log$(N/O).

In the following  sections we will analyse the dependence of abundance slopes and y-intercepts of the abundance radial profile fits on other parent galaxy properties.

\begin{table*}
\centering
\caption{Median value (and corresponding error) for the 12+log(O/H) central abundance (y-intercept) and slope of the radial oxygen abundance gradient in different units for (strongly and weakly) barred and unbarred galaxies separately and for all the galaxies of the sample. N indicates the number of galaxies in each subsample. We include the results obtained using all the different strong-line techniques and calibrations we adopted.}
\label{tab:fits_OH}
\begin{tabular}{lcccccc} 
 \hline
                                &  12 + log(O/H)  &   $\alpha_{O/H}$   &   $\alpha_{O/H}$     &  $\alpha_{O/H}$      &  N  \\ 
                                &  at  R=0            &  (dex kpc$^{-1}$)  &  (dex r$_{25}^{-1}$)  & (dex r$_{e}^{-1}$)   &     \\ 
\hline
\multicolumn{6}{l}{HCM - \citet{epm14}}\\  
\hline
strongly barred               & $8.68\pm0.06$  & $-0.006\pm0.004$  &  $-0.06\pm0.07$  & $-0.03\pm0.02$     & 22\\  %% Revisado el 06/07/20 %% 
weakly   barred               & $8.7\pm0.2$      & $-0.03\pm0.01$      &  $-0.2\pm0.4$      & $ -0.1\pm0.1$         & 9\\
strong + weakly barred  & $8.70\pm0.06$  & $ -0.007\pm0.007$ &  $-0.1\pm0.1$      & $-0.04\pm0.04$     & 31\\
unbarred                         & $8.68\pm0.06$  & $-0.016\pm0.008$  &  $-0.2\pm0.1$      & $-0.08\pm0.06$     & 20\\
all                                   & $8.69\pm0.04$  & $-0.011\pm0.006$  &  $-0.14\pm0.08$  & $-0.05\pm0.04$ & 51\\
\hline
\multicolumn{6}{l}{N2 - \citet{pp04}}\\
\hline
strongly barred             & $8.67\pm0.04$  &$-0.004\pm0.003$  &  $-0.07\pm0.04$  & $-0.03\pm0.01$  &19\\%% Revisado el 06/07/20   %%
weakly   barred             & $8.64\pm0.04$  &$-0.014\pm0.007$  &  $-0.23\pm0.08 $  & $-0.09\pm0.05$  &8\\
strong + weakly barred& $8.66\pm0.03$   &$-0.006\pm0.004$  &  $-0.09\pm0.06$  & $-0.03\pm0.01$  &27\\
unbarred                      & $8.64\pm0.05$   &$-0.020\pm0.009$  &  $-0.26\pm0.07$  & $-0.09\pm0.02$  &17\\
all                                & $8.66\pm0.03$   &$-0.011\pm0.004$  &  $-0.15\pm0.05 $  & $-0.06\pm0.02$  &44\\
\hline
\multicolumn{6}{l}{O3N2 - \citet{pp04}}\\
\hline
strongly barred              & $8.85\pm0.05 $  & $-0.022\pm0.005$  & $-0.3\pm0.1 $   & $-0.09\pm0.04$  & 19\\ %% Revisado el 06/07/20  %% 
weakly   barred              & $8.7\pm0.1$       & $-0.029\pm0.007$  & $-0.4\pm0.2 $   & $-0.2\pm0.1$      & 8\\
strong + weakly barred & $8.79\pm0.07 $  & $-0.023\pm0.005$  & $-0.32\pm0.08$ & $-0.10\pm0.03$  & 27\\
unbarred                       & $8.8\pm0.1   $    & $-0.037\pm0.008$  & $-0.42\pm0.08$ & $-0.20\pm0.04$  & 17\\
all                                 & $8.79\pm0.06 $  & $-0.026\pm0.005$  &$ -0.34\pm0.06$ & $-0.15\pm0.03$  & 44\\ %antes 46 (con CNR)
\hline
\multicolumn{6}{l}{O3N2 - \citet{marino13}}\\
\hline
strongly barred         & $8.61\pm0.04$      & $-0.014\pm0.004$  & $-0.18\pm0.07$ & $-0.06\pm0.02$  & 19\\%% Revisado el 08/07/20   %% 
weakly   barred          & $8.5\pm0.1  $       & $-0.021\pm0.006$  & $-0.2\pm0.1  $ & $-0.14\pm0.08$ & 8\\
strong + weakly barred & $8.59\pm0.05$  & $-0.014\pm0.004$  & $-0.19\pm0.05$ & $-0.08\pm0.02$  & 27\\
unbarred                  & $8.56\pm0.07$       & $-0.025\pm0.005$  & $-0.26\pm0.06$ & $-0.13\pm0.03$  & 17\\
all                            & $8.60\pm0.04$       & $-0.017\pm0.003$  & $-0.22\pm0.04$ & $-0.09\pm0.02$  & 44\\ 
\hline
\multicolumn{6}{l}{N2O2 - \citet{b07}}\\  
\hline
strongly barred              & $8.71\pm0.07 $  & $-0.021\pm0.007$   & $-0.3\pm0.1  $    & $-0.11\pm0.03$ &18\\ %% Revisado el 06/07/20  %% 
weakly   barred              & $8.65\pm0.08$   & $-0.027\pm0.007$   & $-0.4\pm0.1  $    & $-0.2\pm0.1$     &8\\
strong + weakly barred & $8.67\pm0.06$   & $-0.024\pm0.005 $  & $-0.34\pm0.09$   & $-0.12\pm0.03$ &26\\
unbarred                       & $8.6\pm0.1$       & $-0.04\pm0.01$       & $-0.41\pm0.07$   & $-0.17\pm0.03$ &17\\
all                                  & $8.64\pm0.06$  &$ -0.030\pm0.006$   & $-0.38\pm0.05$   & $-0.14\pm0.03$ &43\\  %45 sin quitar CNR
\hline 
\multicolumn{6}{l}{R23 - \citet{m91,kkp99}}\\ 
\hline
strongly barred                &  $9.0\pm0.1$    & $-0.013\pm0.006$ &$-0.2\pm0.1$      & $-0.09\pm0.04$ &  19\\   %% Revisado el 06/07/20 %%  
weakly   barred                &  $9.0\pm0.1$    & $ -0.030\pm0.008$ &$-0.4\pm0.2 $    & $-0.2\pm0.1$ &  9\\
strong + weakly barre d  &  $8.98\pm0.08$ & $-0.019\pm0.006$ &$-0.25\pm0.09$  & $-0.11\pm0.04$ &  28\\
unbarred                         &  $8.97\pm0.09$ & $-0.035\pm0.009$ &$-0.37\pm0.06$  & $-0.16\pm0.03$ &  20\\
all                                   &  $8.97\pm0.06$ & $-0.022\pm0.006$ &$-0.32\pm0.05$  & $-0.14\pm0.03$ &  48\\
%\multicolumn{6}{l}{R23 - \citet{Zaritsky94}}\\ 
%\hline
%strongly barred             &  $9.1\pm0.1$    &  $-0.016\pm0.008$  & $-0.2\pm0.1 $    & $-0.11\pm0.05$ & 20\\ %% Actualizado el 04/11/19 (*****)
%weakly   barred             &  $9.2\pm0.2$    &  $-0.04\pm0.01$      & $-0.5\pm0.2 $    & $-0.29\pm0.07$ &  8\\
%strong + weakly barred&  $9.2\pm0.1$    &  $-0.024\pm0.008$  & $-0.4\pm0.1 $    & $-0.14\pm0.05$ & 28\\
%unbarred                       &  $9.1\pm0.1$    &  $-0.05\pm0.01$      & $-0.48\pm0.07 $& $-0.20\pm0.04$ & 20\\
%all                                &  $9.14\pm0.07$  &  $-0.028\pm0.006 $ & $-0.45\pm0.07 $& $-0.17\pm0.04$ & 48\\%% 
%\hline
\hline
\multicolumn{6}{l}{$R$ calibration - \citet{PilyuginGrebel2016}}\\
\hline 
 %                                   &  at  R=0                &  (dex kpc$^{-1}$)  &  (dex r$_{25}^{-1}$)  & (dex r$_{e}^{-1}$)   &     \\ 
 strongly barred            & $8.66\pm0.04$      & $-0.010\pm0.006$   & $-0.11\pm0.05$  & $-0.04\pm0.02$  & 17\\%% Revisado el 06/07/20   %% 
weakly   barred             & $8.67\pm0.06$      & $-0.026\pm0.009$   & $-0.3\pm0.1  $    & $-0.15\pm0.07$  & 8  \\
strong + weakly barred & $8.66\pm0.03$      & $-0.013\pm0.006$  & $-0.20\pm0.08$  & $-0.07\pm0.03$  & 25\\
unbarred                       & $8.61\pm0.07$       & $-0.025\pm0.009$  & $-0.29\pm0.07$  & $-0.14\pm0.03$  & 17\\
all                                 & $8.65\pm0.04$       & $-0.020\pm0.005$  & $-0.24\pm0.06$  & $-0.09\pm0.03$  & 42\\ 
\hline\end{tabular}
\end{table*}

%%%%%%%%%%%%%%%%%%%%%%%%%%%%%%%%%%%%%%%%%%%%%%%%%%%%%%%%%%%%%%%%%%%%%%%%%%%%%%%%%%%%%%%%%%%%%%%%%%%%%%%%%%%%%%
%%%%%%%%%%%%%%%%%%%%%%%%%%%%%%%%%%%%%%%%%%%%%%%%%%%%%%%%%%%%%%%%%%%%%%%%%%%%%%%%%%%%%%%%%%%%%%%%%%%%%%%%%%%%%%
%%%%%%%%%%%%%%%%%%%%%%%%%%%%%%%%%%%%%%%%%%%%%%%%%%%%%%%%%%%%%%%%%%%%%%%%%%%%%%%%%%%%%%%%%%%%%%%%%%%%%%%%%%%%%%
\subsection{Intercepts and slopes vs. M$_B$ and T}
\label{slope_vs_M}
The possible connection between radial abundance gradients in spirals and their galaxy mass is of great relevance for evolutionary processes. The gradient was found to depend on galaxy mass (or luminosity) \citep[e.g.][]{tremonti,Ho2015, Bresolin2019} and morphological type \citep[e.g.][]{vila-costas92,Oey93}, with more massive and luminous galaxies (or earlier types) showing a shallower abundance gradient when expressed in dex~kpc$^{-1}$. Recent observational IFU-based data on large data samples yield however contradictory results: \citet{Belfiore2017} find that low-mass galaxies have shallower radial oxygen abundance  gradients than massive ones, while \citet{Sanchez-Menguiano2016} do not detect that trend. \citet{pilyugin2014} do not find a correlation  between N/O and O/H radial gradients in  dex~kpc$^{-1}$ and morphological types, for a galaxy sample of 130 nearby spirals with data based on published spectra from different authors and instrumentation.
%In this section we inspect these relations for our galaxy sample, separately for barred and unbarred galaxies.

The distributions of absolute magnitudes (M$_B$, as a proxy for stellar mass) and  morphological types (T-type) for our galaxies are similar for our sub-samples of barred and unbarred galaxies (see fig.~2 in \citetalias{paperI}), and therefore the differences in gradient slopes that we observed earlier can not be due on  first order to biases in these parameters. It  is worth  checking now whether abundance gradients in barred and unbarred galaxies show the same trends with M$_B$ and T-type. 

%%%%%%%%%%%%%%%%%%%%%%%%%%%%%%%%%%%%%%%%%%%%%%%%%%%%%%%%%%%%%%%%%%%%%%%%%%%%%%%%%%%%%%%%%%%%%%%%%%%%%%%%%%%%%%
%%%%%%%%%%%%%%%%%%%%%%%%%%%%%%%%%%%%%%%%%%%%%%%%%%%%%%%%%%%%%%%%%%%%%%%%%%%%%%%%%%%%%%%%%%%%%%%%%%%%%%%%%%%%%%
%%%%%%%%%%%%%%%%%%%%%%%%%%%%%%%%%%%%%%%%%%%%%%%%%%%%%%%%%%%%%%%%%%%%%%%%%%%%%%%%%%%%%%%%%%%%%%%%%%%%%%%%%%%%%%
\subsubsection{Intercepts and slopes vs. M$_B$}
\label{fits_vs_MB}
\begin{figure*}
\vspace{-0.3cm}\includegraphics[width=0.64\textwidth,clip,trim = 0cm 0.3cm 0cm 0.2cm]{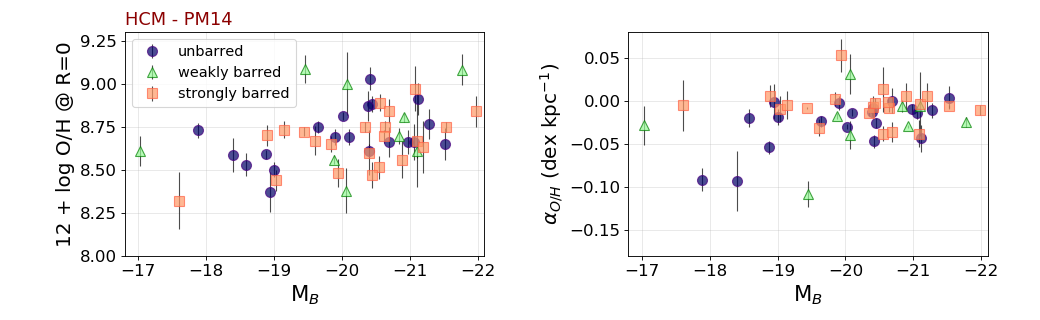}
\vspace{-0.1cm}\includegraphics[width=0.64\textwidth,clip,trim = 0cm 0.2cm 0cm 0.2cm]{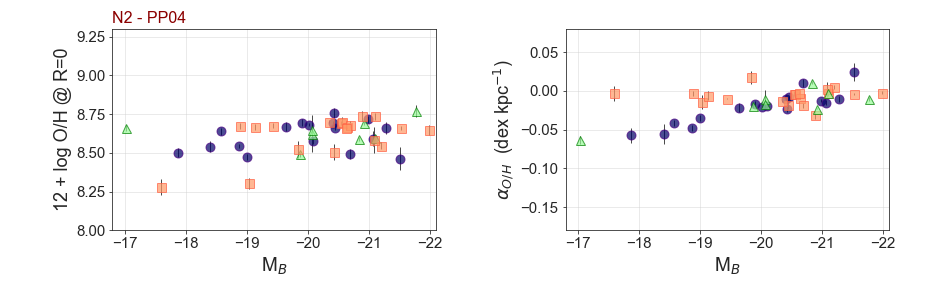}
\vspace{-0.1cm}\includegraphics[width=0.64\textwidth,clip,trim = 0cm 0.2cm 0cm 0.2cm]{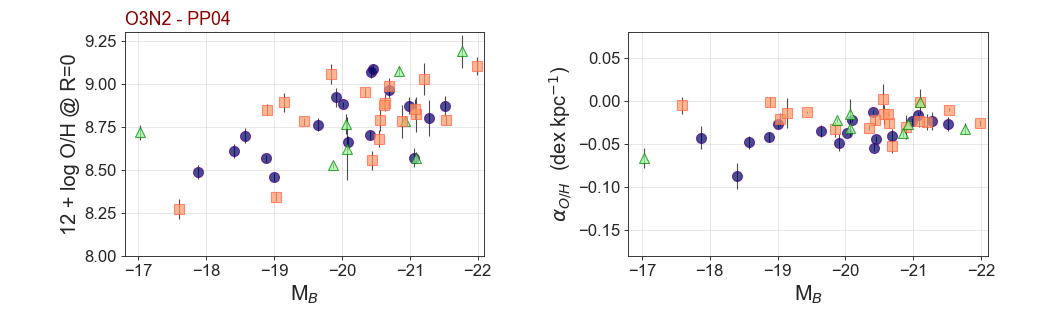}
\vspace{-0.1cm}\includegraphics[width=0.64\textwidth,clip,trim = 0cm 0.2cm 0cm 0.2cm]{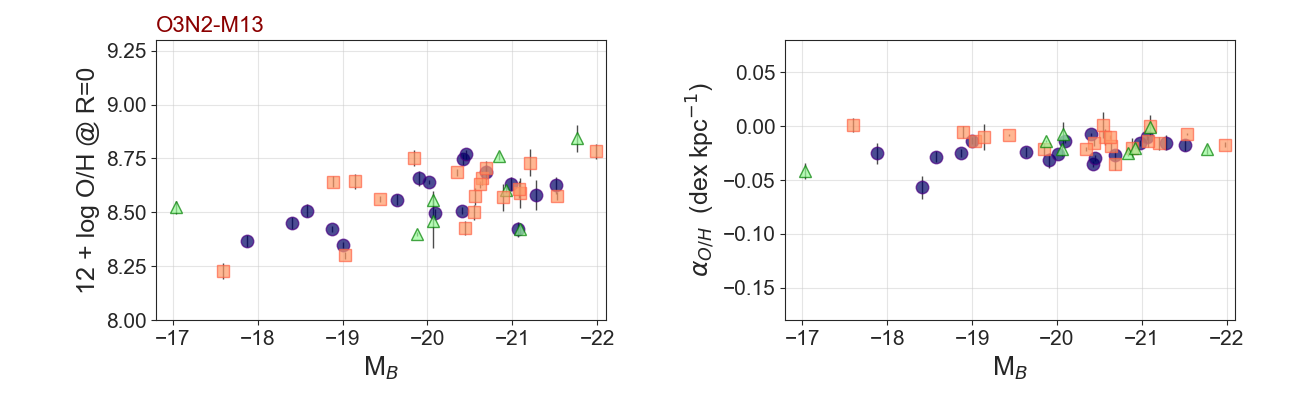}
\vspace{-0.1cm}\includegraphics[width=0.64\textwidth,clip,trim = 0cm 0.2cm 0cm 0.2cm]{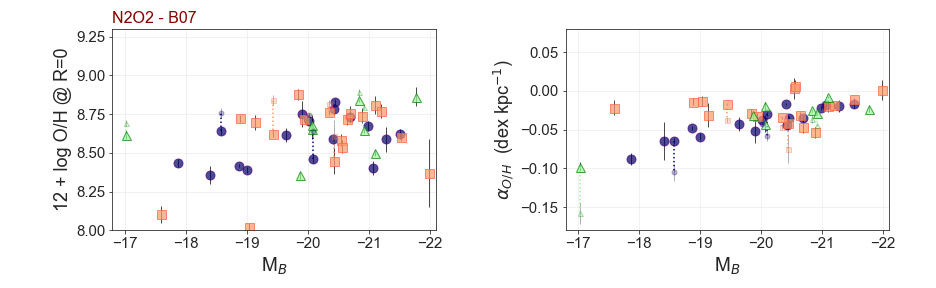}
\vspace{-0.1cm} \includegraphics[width=0.64\textwidth,clip,trim = 0cm 0.2cm 0cm 0.2cm]{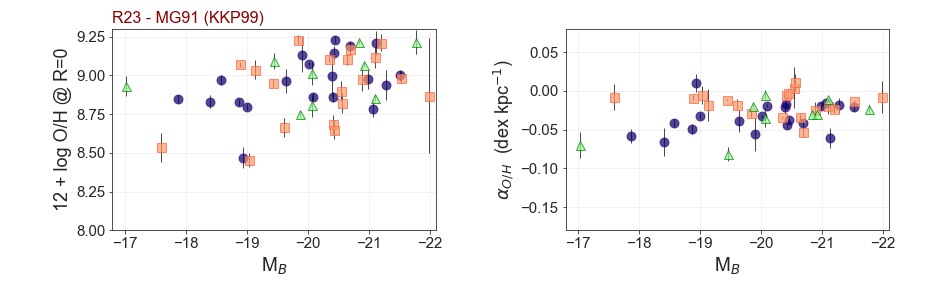}
\vspace{-0.1cm} \includegraphics[width=0.64\textwidth,clip,trim = 0cm 0.2cm 0cm 0.2cm]{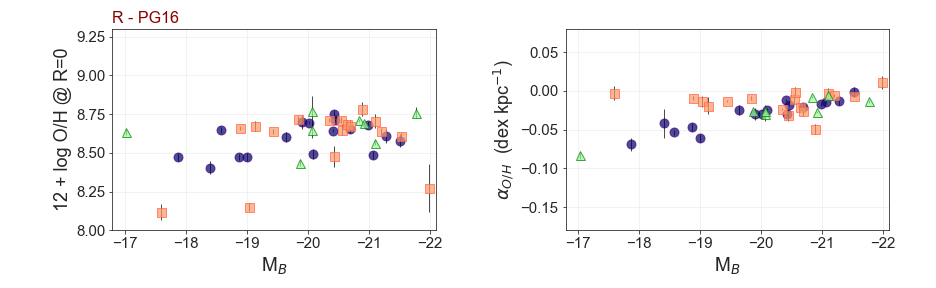}
 \caption{Central oxygen abundance (y-intercept from radial fits, {\em left}) and slope of the oxygen abundance radial gradient in  dex~kpc$^{-1}$ ({\em right}) for different strong line 12+$\log$(O/H) diagnostics. Unbarred galaxies are shown  with blue circles, weakly barred galaxies with green triangles and strongly barred galaxies with orange squares. The small data points in the panel for N2O2 correspond to the inner slopes and y-intercepts for galaxies in which  a double linear fit is performed to the radial abundance profile (see Sect.~\ref{breaks}).
  \label{empi_vs_M}}
\end{figure*}

%We first analyse the central abundance and the abundance gradients as a function of the absolute B-band magnitude, M$_B$.
M$_B$ has been used in the past as a proxy for stellar mass when the latter  is not available. It is important however, to bear in mind that M$_B$ is not a reliable indicator of stellar mass, as it can be affected by dust extinction and current star formation. The left-hand panels in Fig.~\ref{empi_vs_M} show the relationship between  the central O/H abundances (y-intercept) and M$_B$, and the right-hand ones show the $\log$(O/H) gradient slope in dex~kpc$^{-1}$ as a function of M$_B$, for the different strong-line oxygen  abundance diagnostics. Unbarred, weakly and strongly barred galaxies are shown with different colours and symbols. A first look at the central abundances shows a general, and expected, trend for 12+$\log$~(O/H) to increase in more luminous galaxies, but the strength and the slope of this relation is highly dependent on the metallicity diagnostic employed, with abundances obtained with N2 and O3N2 (as calibrated by M13) showing little dependence with M$_B$, while the luminosity\,-\,metallicity  relation is more evident with the O3N2 diagnostic (as calibrated by PP04). No clear difference is observed between strongly barred  and unbarred galaxies.

The metallicity gradient\,-\,M$_B$ relation does show a clearly different behaviour for strongly barred and unbarred galaxies: strongly barred galaxies show a rather shallow oxygen abundance gradient across the whole luminosity range and for all the strong line diagnostics. However, the metallicity gradient  in unbarred galaxies steepens as the luminosity decreases (for slopes in dex~kpc$^{-1}$). This trend is seen in the gradients obtained using each diagnostic, but it is clearer with N2O2 and N2. The same plots show that luminous (M$_B \lesssim-19.5$)  galaxies show shallow gradients, regardless of whether they have a bar or not.
%\boldgreen{}
We are aware that the number of low luminosity barred galaxies in our sample is low. In fact there is only one barred galaxy with M$_B \gtrsim-18.5$, NGC~4395. We analysed the gradient of this galaxy carefully in \citetalias{paperI}, and showed that the metallicity gradient is shallow with all strong-line methods, yielding an average value of $\alpha_{O/H}=-0.005\pm0.007$~dex~kpc$^{-1}$, that is considerably distant from the range of values covered by unbarred galaxies of similar  M$_B$, that goes from $\sim~-0.06$ to $-0.10$ dex~kpc$^{-1}$  (ignoring methods based on the O3N2 indicator, in which the range of observed slopes is much smaller). Although it would be  highly desirable to enlarge the number statistics at low luminosity to confirm this, our results point to a rather constant metallicity slope for barred galaxies across the whole luminosity range.

\begin{figure*}
\includegraphics[width=0.72\textwidth]{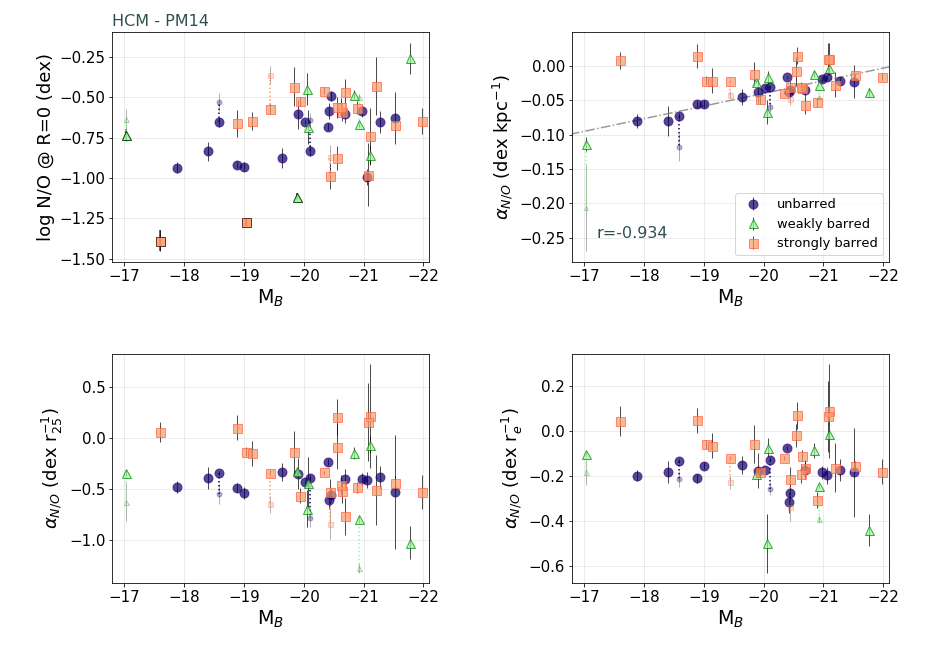}
\caption{Central N/O abundance  (y-intercept from radial profile fits, {\em top-left}) and slope of the fit to the  $\log$(N/O) radial abundance profile, as derived from HCM, as a function of the $B$-band absolute magnitude of the galaxies, for slopes in dex~kpc$^{-1}$ ({\em top-right}), dex~r$_{25}^{-1}$ ({\em bottom-left}) and  dex~r$_{e}^{-1}$ ({\em bottom-right}). The grey straight dashed line shows the linear fit to the slope in dex~kpc$^{-1}$ as a function of M$_B$ for unbarred galaxies alone. See Section~\ref{slope_vs_M} for details. Symbols as in Fig.~\ref{empi_vs_M}.\label{NO_vs_M}}
%\end{figure*}
%\begin{figure*}
\includegraphics[width=0.72\textwidth]{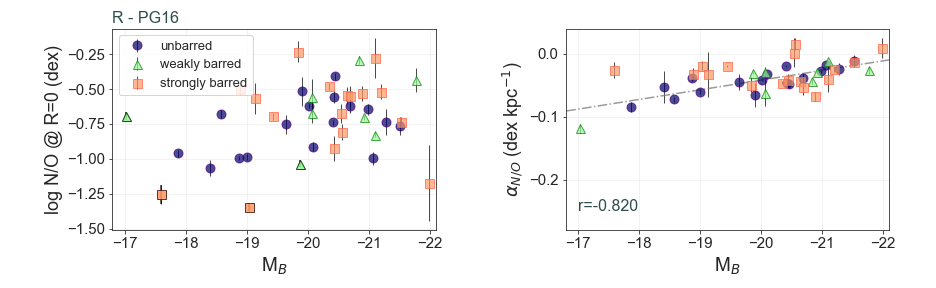}\\
\includegraphics[width=0.72\textwidth]{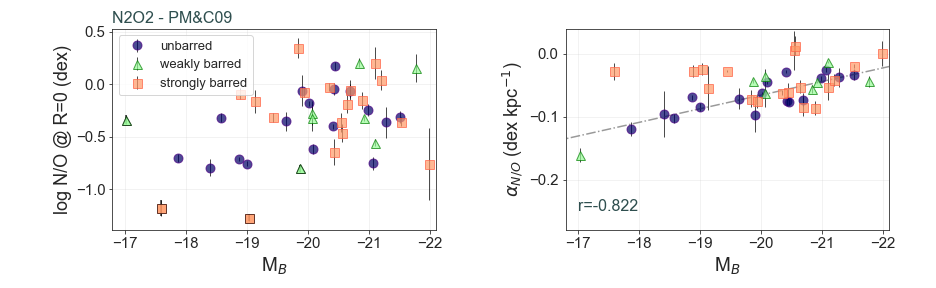}\\
\includegraphics[width=0.72\textwidth]{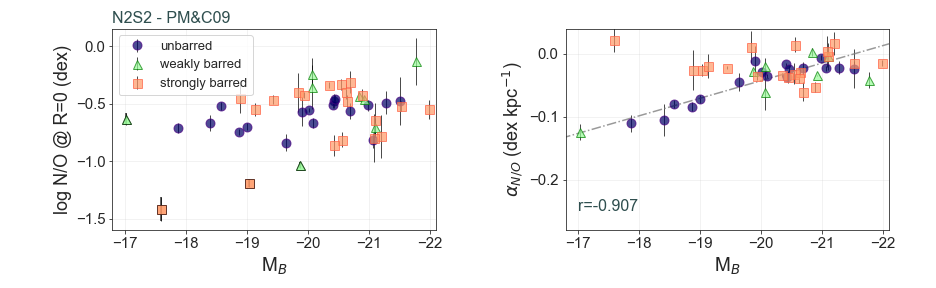}
\caption{Central N/O abundance  (y-intercept from radial profile fits, {\em left}) and slope of the fit to the  $\log$(N/O) radial abundance profile  ({\em right}) as a function of the $B$-band absolute magnitude of the galaxies for N/O abundances derived from the $R$ calibration, N2O2 and N2S2 as indicated in the plots. The grey straight dashed line shows the linear fit to the slope in dex~kpc$^{-1}$ as a function of M$_B$ for unbarred galaxies alone. Symbols as in Fig.~\ref{empi_vs_M}.\label{NO_vs_M_otros}}
\end{figure*}

The relation between the abundance gradient and M$_B$ for unbarred galaxies is stronger for $\log$(N/O). This can be seen in Fig.~\ref{NO_vs_M} ({\em top-right}) for HCM and in the right-hand panels in Fig.~\ref{NO_vs_M_otros} for the $R$ calibration, N2O2 and N2S2, for the slopes in dex~kpc$^{-1}$. A linear fit to the unbarred galaxies data points for HCM yields a correlation coefficient r=$-0.93$ for the following relation:
\begin{equation}
\alpha_{N/O} ~ {\mbox (dex~kpc^{-1})} = -(0.41\pm0.04) - (0.018\pm0.002)\times \mbox{M}_B
\end{equation}
The linear fit to the same relation as derived from the $R$ calibration also yields a high correlation coefficient, r=$-0.82$, with a slope and y-intercept of $-(0.35\pm0.06)$ and  $-(0.015\pm0.003)$, respectively, in good agreement, within errors, with the values derived for HCM. For N2O2 and N2S2 $\alpha_{N/O}$ is also strongly correlated with M$_B$ for unbarred galaxies (r=$-0.82$ and $-0.91$, respectively), but with a steeper slope ($\sim-0.025$) and a lower y-intercept, possibly because these calibrations do not match the $T_e$-based scale as closely as HCM and the $R$ calibration for our \hii\ region sample (\citetalias{paperI}).

However, strongly barred galaxies have a shallow abundance profile gradient in the whole luminosity range, with an average value of $-0.021\pm0.009$~ dex~kpc$^{-1}$ with HCM, and  $-0.032\pm0.009$~ dex~kpc$^{-1}$ from the $R$ calibration. The  rather uniform (and shallow) slope  values for strongly barred galaxies and the observed tendency for unbarred spirals to have shallower gradients as their total luminosity increases imply  that bright (M$_B\lesssim-19.5$)  barred and unbarred galaxies are indistinguishable in terms of their slopes, the two  sets of galaxies presenting rather flat O/H and N/O abundance profiles.

The normalisation of $\alpha_{N/O}$ by either r$_{25}$ or r$_e$ (bottom panels in Fig.~\ref{NO_vs_M}) leads to a rather constant value for unbarred galaxies, $\alpha_{N/O}=-0.18$~dex~r$_e^{-1}$, with both  with HCM and the $R$ calibration, while barred  galaxies show a larger spread of values (scatter $\sim$0.11-0.13 ~dex~r$_e^{-1}$, c.f. $\sim$0.05-0.08~dex~r$_e^{-1}$ for unbarred galaxies). This result supports the idea of a {\em common abundance gradient} \citep[][]{sanchez14,BresolinKennicutt2015} when the slopes are expressed in dex~r$_e^{-1}$, that our plot suggests is better defined (because of the smaller scatter) by the unbarred galaxies. The same constancy in slopes, when normalised to  r$_e^{-1}$, is observed for $12+\log$~(O/H) with all the strong-line methods employed here, with a larger spread of values for barred galaxies.

The relationship between the central N/O abundances (y-intercept)  and  M$_B$  is shown in the  top-left panel in  Fig.~\ref{NO_vs_M} for HCM, and in the left-hand panels in Fig.~\ref{NO_vs_M_otros} for the rest of the methods. In spite of the large scatter, there is a general trend for $\log$~(N/O) to increase in more luminous galaxies. There is no clear  difference between strongly barred  and unbarred galaxies. However, if we ignore Magellanic-type barred galaxies (marked with grey edge symbols and showing the lowest log(N/O) values), there seems to be a trend for lower luminosity barred galaxies to have larger central log(N/O). 

%{\em Add comment on comparison or our mena values for unbarred galaxies with data on Table~\ref{tab:universal_gradient}.}
%% Garnett & Shields 1987, ya mestran dependencia entre O/H con masa de la galaxia.

%%%%%%%%%%%%%%%%%%%%%%%%%%%%%%%%%%%%%%%%%%%%%%%%%%%%%%%%%%%%%%%%%%%%%%%%%%%%%%%%%%%%%%%%%%%%%%%%%%%%%%%%%%%%%%
%%%%%%%%%%%%%%%%%%%%%%%%%%%%%%%%%%%%%%%%%%%%%%%%%%%%%%%%%%%%%%%%%%%%%%%%%%%%%%%%%%%%%%%%%%%%%%%%%%%%%%%%%%%%%%
%%%%%%%%%%%%%%%%%%%%%%%%%%%%%%%%%%%%%%%%%%%%%%%%%%%%%%%%%%%%%%%%%%%%%%%%%%%%%%%%%%%%%%%%%%%%%%%%%%%%%%%%%%%%%%
\subsubsection{Intercepts and slopes vs. morphological T-type}
\begin{figure*}
\vspace{-0.2cm}\includegraphics[width=0.64\textwidth,clip,trim = 0cm 0.2cm 0cm 0.2cm]{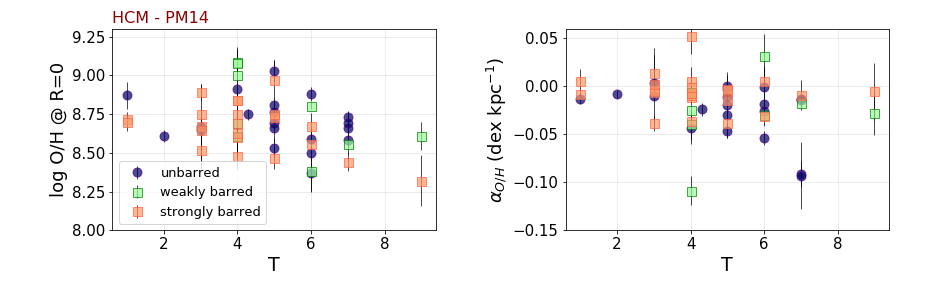}
\vspace{-0.1cm}\includegraphics[width=0.64\textwidth, clip,trim = 0cm 0.2cm 0cm 0.2cm]{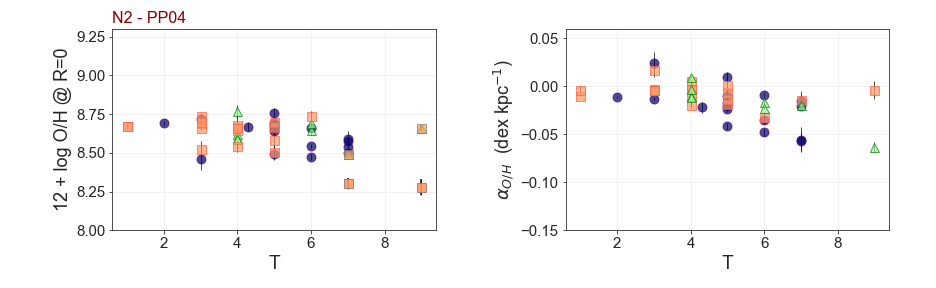}
\vspace{-0.1cm}\includegraphics[width=0.64\textwidth, clip,trim = 0cm 0.2cm 0cm 0.2cm]{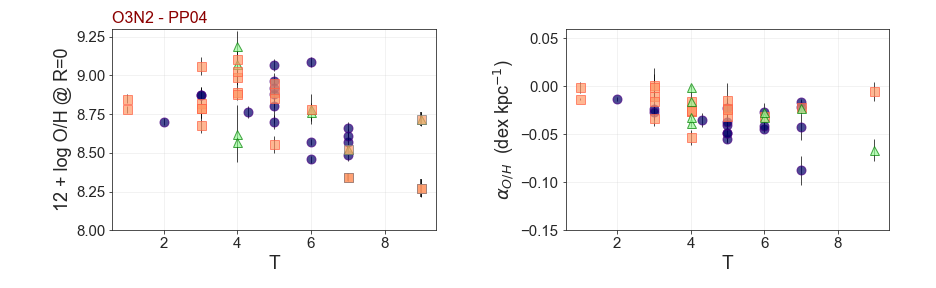}
\vspace{-0.1cm} \includegraphics[width=0.64\textwidth,clip,trim = 0cm 0.2cm 0cm 0.2cm]{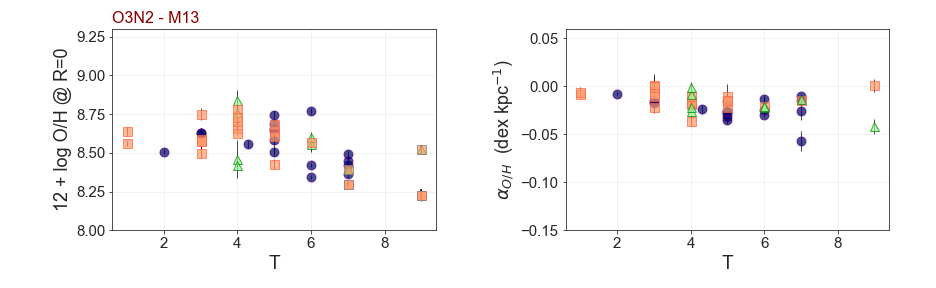}
\vspace{-0.1cm}\includegraphics[width=0.64\textwidth,clip,trim = 0cm 0.2cm 0cm 0.2cm]{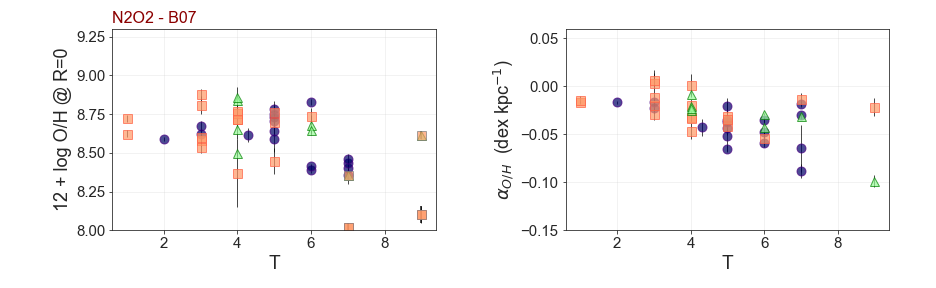}
\vspace{-0.1cm} \includegraphics[width=0.64\textwidth,clip,trim = 0cm 0.2cm 0cm 0.2cm]{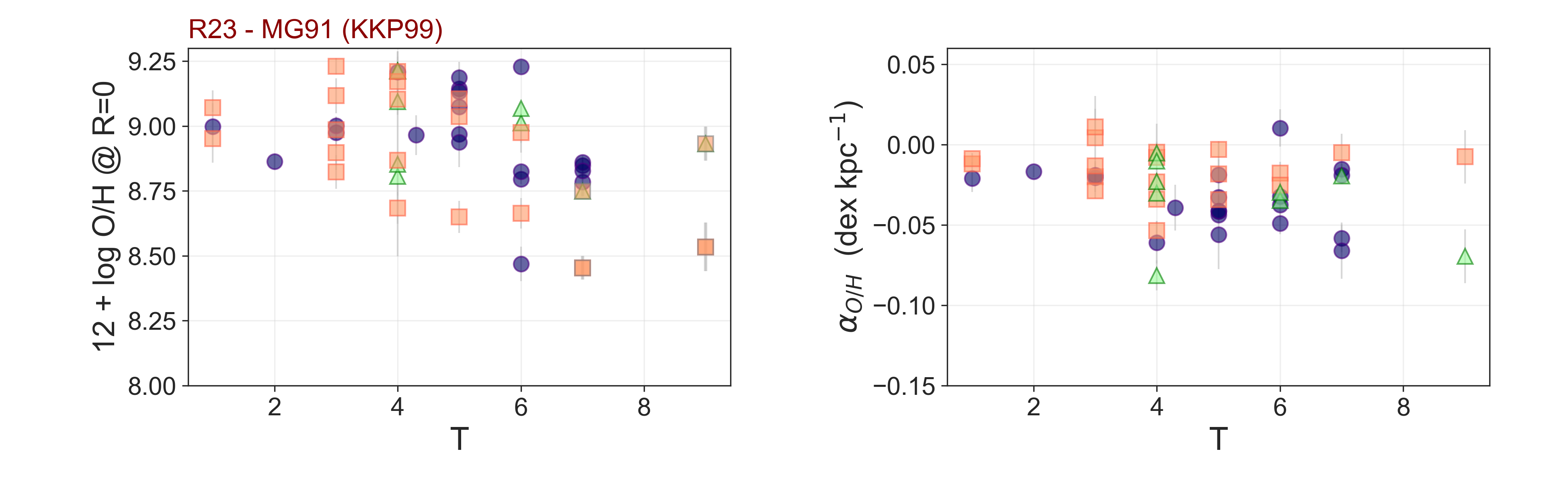}
\vspace{-0.1cm}\includegraphics[width=0.64\textwidth,clip,trim = 0cm 0.2cm 0cm 0.2cm]{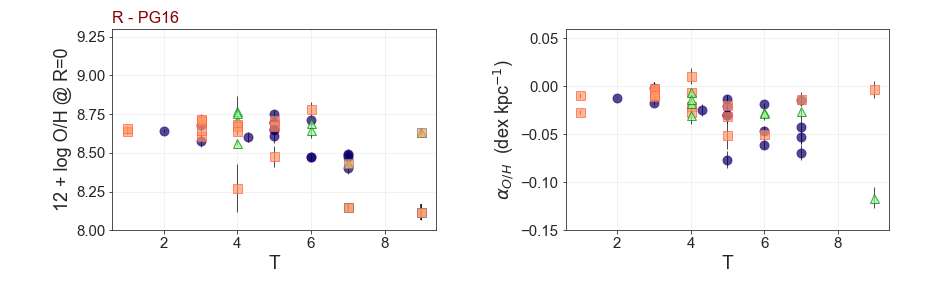}
 \caption{Central oxygen abundance (y-intercept from radial fits, {\em left}), and slope of the oxygen abundance radial gradient in  dex~kpc$^{-1}$ ({\em right}) for different strong line 12+$\log$(O/H) diagnostics as a function of the morphological T-type.  Symbols as in Fig.~\ref{empi_vs_M}. \label{OH_vs_T}}
\end{figure*}

Figs.~\ref{OH_vs_T} shows the central oxygen abundance ({\em left-hand} panels) and the slope $\alpha_{O/H}$ ({\em right-hand} panels) as a function of the morphological T-type. The dispersion in the plots is high, but the diagrams  show: (a) a tendency for galaxies of earlier T-types to  have higher central oxygen abundances,   (b) earlier type galaxies have, on average, shallower abundance gradients, and (c) for later T-types there is a considerable spread in gradient values  (in dex~kpc\me). No difference between barred and unbarred galaxies is seen in central $\log$(O/H), as we already found in the histograms (Fig.~\ref{histog_fits_HCM_OH} and \ref{histog_empi}). When inspecting more carefully the right panel of Fig.~\ref{OH_vs_T} we can see that the trend for later type galaxies to have steeper gradients  is driven by unbarred (and weakly barred) galaxies. In other words, in our galaxy sample shallow gradients can be found for all T-types, but the steepest gradients are seen only in late-type unbarred (or weakly barred) galaxies. 

\begin{figure*}
\hspace{-0.5cm}\includegraphics[width=0.75\textwidth]{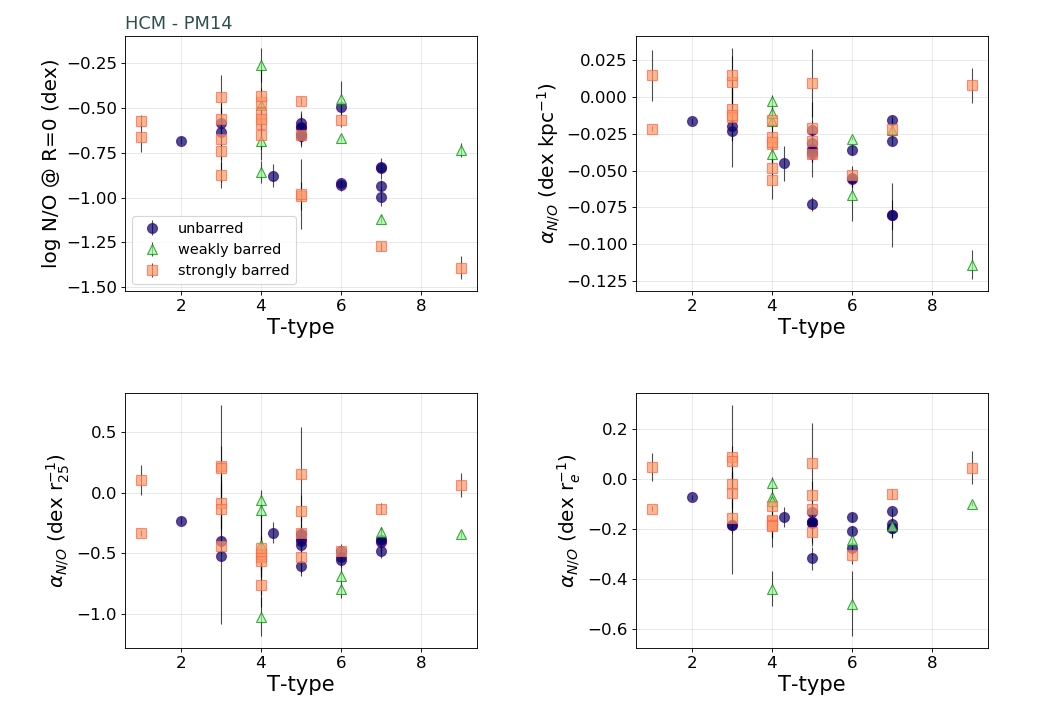}
 \caption{Central N/O abundance (y-intercept from radial fits, {\em left}), and  gradient slope of the $\log$(N/O)  radial profile as derived from HCM (with different normalizations) as a function of the morphological T-type. Symbols as in Fig.~\ref{empi_vs_M}. \label{NO_vs_T}}
\end{figure*}

The central abundance and the slope for the $\log$(N/O) radial abundance profile (Fig.~\ref{NO_vs_T}, for HCM) show a similar trend with T-type to the one observed for $\log$(O/H), with later types showing  on average lower central N/O abundances and a larger spread on $\alpha_{N/O}$ values (in dex~kpc$^{-1}$). No clear difference is seen between strongly barred and unbarred galaxies in terms of their dependence on  T, apart from a similar behaviour as observed for $\alpha_{O/H}$: the larger spread towards later types seems to be caused by unbarred and weakly barred galaxies. 

The $\alpha_{N/O}$ and $\alpha_{O/H}$ slopes, normalised to r$_{25}$ or r$_e$, show the same trend for a common value with T-types for unbarred galaxies, with a larger spread for barred galaxies, as we already mentioned in the previous section. 
%% Perez-Montero2016 tambien encuentra esa tendencia con tipos

%\begin{minipage}
%\end{minipage}

%trim={<left> <lower> <right> <upper>}
%[width=4.5cm,clip,trim = 5cm 0.cm 2.5cm 0.5cm,angle=-0]{

%%%%%%%%%%%%%%%%%%%%%%%%%%%%%%%%%%%%%%%%%%%%%%%%%%%%%%%%%%%%%%%%%%%%%%%%%%%%%%%%%%%%%%%%%%%%%%%%%%%%%%%%%%%%%%
%%%%%%%%%%%%%%%%%%%%%%%%%%%%%%%%%%%%%%%%%%%%%%%%%%%%%%%%%%%%%%%%%%%%%%%%%%%%%%%%%%%%%%%%%%%%%%%%%%%%%%%%%%%%%%
%%%%%%%%%%%%%%%%%%%%%%%%%%%%%%%%%%%%%%%%%%%%%%%%%%%%%%%%%%%%%%%%%%%%%%%%%%%%%%%%%%%%%%%%%%%%%%%%%%%%%%%%%%%%%%
\section{Dependence with bar parameters} 
\label{bar_params}
Bar-induced gas flows are a strong function of the {\em bar strength}, a parameter that quantifies the gravitational effect of the stellar bar on the surrounding disc. Stronger bars are predicted to supply gas towards the galaxy centres at a higher rate \citep[e.g.][]{athanassoula,ReganTeuben,KimStone2012}. The bar ellipticity is a proxy for the bar strength \citep[e.g.][]{abraham00,Eskridge2004,SimonDiaz2016}.  We explore here the relation between  intercepts and slopes of the radial abundance profiles with the bar parameters.

%If radial abundance profiles are flattened by the bar induced mixing of gas of different content in heavy elements, a relation between slope and bar parameters could in principle be expected.

\begin{figure*}
\includegraphics[width=0.8\textwidth]{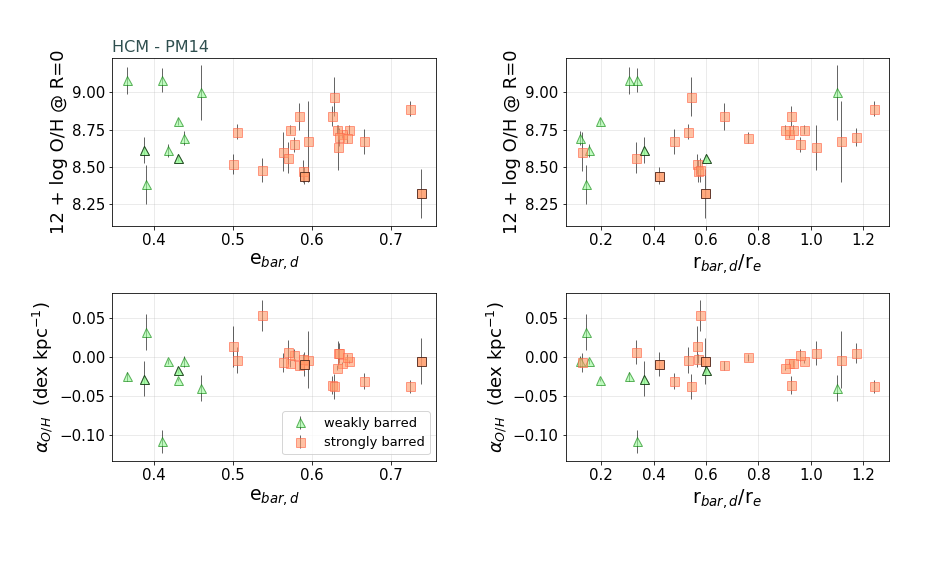}  %logOH_fits_vs_bar_params_ellip_SF.png
\vspace{-1cm}
\includegraphics[width=0.8\textwidth]{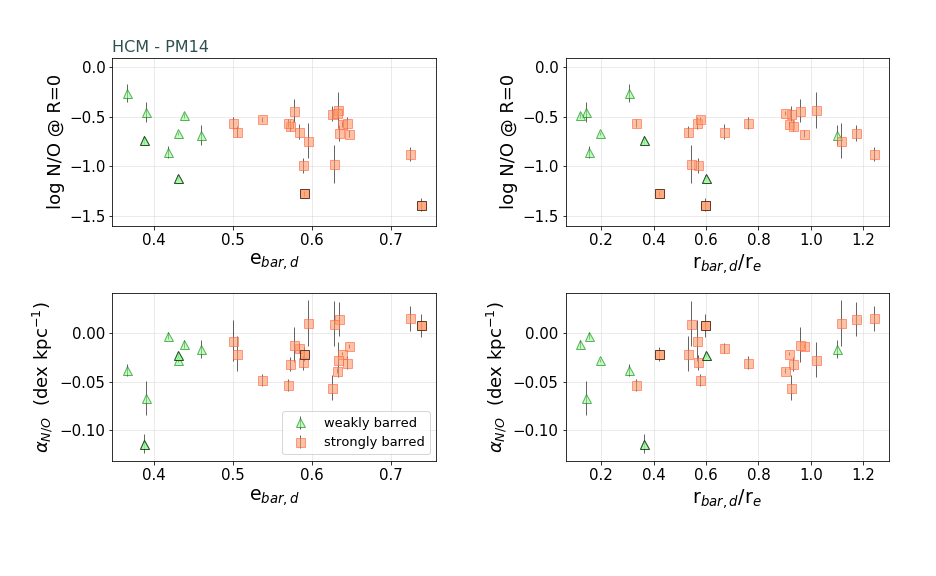}  %logNO_fits_vs_bar_params_ellip_SF.png
\caption{{\em Top:} Central (y-intercept) and slope for the $12+\log$~(O/H) radial abundance profile (from HCM) as a function of the deprojected bar ellipticity (left panels) and bar radius normalised to $r_e$ (right panels). {\em Bottom:} Same as in the four top panels but for $\log$(N/O). The later-type galaxies of the sample (NGC~925, NGC~1313, NGC~4395 and NGC~4625) are indicated with grey edge symbols. \label{HCM_vs_bar_params}}
\end{figure*}

Fig.~\ref{HCM_vs_bar_params} shows the central abundances and slopes (in dex~kpc\me) for $12+\log$~(O/H), in the top panels, and $\log$(N/O), bottom panels, as a function of the deprojected bar ellipticity (left) and the bar radius normalised to the disc effective radius of the galaxy (right). The abundances shown in the figure have been estimated with the HCM method, but the conclusions stated  below can also be drawn for the remaining  diagnostics. We do not observe any clear trend between these parameters,
%%%%%%CHECK two for
apart maybe for a tendency for a larger  dispersion in slope for the galaxies with shorter ($r_{bar,d}/r_e \lesssim0.6$) and less elliptical (or weaker) bars ($e_{bar,d}\lesssim0.5$) for $\log$(N/O).

Galaxies with long bars  ($r_{bar,d}/r_e \gtrsim 0.6$) exhibit rather constant central $\log$(N/O) and   $12+\log$(O/H) values, whereas galaxies with shorter bars show a wider range of central abundances. This is not seen in the central abundances {\em vs.} bar ellipticity plots, in part due to the lower central abundance values (especially in N/O) for the later-type galaxies of the sample (NGC~925, NGC~1313, NGC~4395 and NGC~4625), indicated with grey edge symbols. The lack of a relation between central abundances and bar strength (or ellipticity) is in agreement with previous studies (see e.g. \citet{bulbos,ellison,chapelon99,considere,cacho} for  O/H,  and  \citet{bulbos}, for N/O). However, our results disagree with the findings of \citet{Martin94} and the predictions by \cite{friedli_benz1995} for the existence of a correlation between the $\log$~(O/H) radial gradient slope (in dex~kpc\me) and the bar axis ratios (implying shallower gradients in galaxies with the stronger bars).% although other authors have not found these relations either \citep{considere}.

It is worth pointing out that the lack of a relationship between the bar properties and the abundance gradient (and/or central abundances) means that the current length or strength of the bar is not related to current properties of the  radial abundance distribution. However, it does not rule out the bar as a relevant agent or contributor to re-shape radial abundance gradients, since both the gradients and the bar parameters evolve with time.
%during their evolution bars change both length and strength \citep[e.g.][]{athanassoula13,cole}.

In summary, if the radial abundance profile flattening observed is due to the mixing produced by the bars, it seems that any bar, provided that it has considerable ellipticity ($e_{bar,d}\gtrsim0.5$) or length ($r_{bar,d}/r_e \gtrsim 0.6$), is able to flatten the abundance profile, with no dependence of the observed gradient on the current bar properties.
%% Puede ser util Section 6.1 de Friedli & Benz 1995, sobre relacion gradiente con b/a de la barra.
%% Martin & Roy 1995, encuentran una dependencia de pendiente con b/a de la barra

%galaxy & R$_{bar}$/R$_e$ & m$_s$ & em$_s$ & b$_s$ & eb$_s$ & N & r & $\chi^2_s$ & $\sigma_s$ & m$_{in}$ & em$_{in}$ & b$_{in}$ & eb$_{in}$ & m$_{ou}$ & em$_{ou}$ & b$_{ou}$ & eb$_{ou}$ & r$_{br}$ & er$_{br}$ &  $\chi^2_d$ & $\sigma_d$ & m$_d$ & em$_d$ & b$_d$ & eb$_d$ & N$_d$ & r$_d$ & discarded \\
%galaxy  & m$_s$ & em$_s$ & b$_s$ & eb$_s$ & N & r & $\chi^2_s$ & $\sigma_s$ & m$_{in}$ & em$_{in}$ & b$_{in}$ & eb$_{in}$ & m$_{ou}$ & em$_{ou}$ & b$_{ou}$ & eb$_{ou}$ & r$_{br}$ & er$_{br}$ &  $\chi^2_d$ & $\sigma_d$ & m$_d$ & em$_d$ & b$_d$ & eb$_d$ & N$_d$ & r$_d$  \\
%%%% NOTA: Quito dos mega tablas con los resultados de los ajustes. Están incluidas
%%%% en NO_bars_v3.tex. Se crearon con los scripts:
%%
%% Table created with 'profile_stat_v1.ipynb' - 'fits_o.tex' 
%% Table created with 'profile_stat_v1.ipynb' - 'fits_o.tex' 

%%%%%%%%%%%%%%%%%%%%%%%%%%%%%%%%%%%%%%%%%%%%%%%%%%%%%%%%%%%%%%%%%%%%%%%%%%%%%%%%%%%%%%%%%%%%%%%%%%%%%%%%%%%%%%
%%%%%%%%%%%%%%%%%%%%%%%%%%%%%%%%%%%%%%%%%%%%%%%%%%%%%%%%%%%%%%%%%%%%%%%%%%%%%%%%%%%%%%%%%%%%%%%%%%%%%%%%%%%%%%
%%%%%%%%%%%%%%%%%%%%%%%%%%%%%%%%%%%%%%%%%%%%%%%%%%%%%%%%%%%%%%%%%%%%%%%%%%%%%%%%%%%%%%%%%%%%%%%%%%%%%%%%%%%%%%
\section{Further effects on profile fits}
\label{additional_considerations}
Before starting with the discussion of the results presented in previous sections, we want to comment on two effects that could affect the radial abundance profiles of galaxies and therefore could have an influence on the derived conclusions. These are the presence of radial abundance profile {\em breaks} in the galaxies and the potentially different chemical abundance properties of \hii\ regions located within galactic bars. 

%%%%%%%%%%%%%%%%%%%%%%%%%%%%%%%%%%%%%%%%%%%%%%%%%%%%%%%%%%%%%%%%%%%%%%%%%%%%%%%%%%%%%%%%%%%%%%%%%%%%%%%%%%%%%%
%%%%%%%%%%%%%%%%%%%%%%%%%%%%%%%%%%%%%%%%%%%%%%%%%%%%%%%%%%%%%%%%%%%%%%%%%%%%%%%%%%%%%%%%%%%%%%%%%%%%%%%%%%%%%%
%%%%%%%%%%%%%%%%%%%%%%%%%%%%%%%%%%%%%%%%%%%%%%%%%%%%%%%%%%%%%%%%%%%%%%%%%%%%%%%%%%%%%%%%%%%%%%%%%%%%%%%%%%%%%%
\subsection{Radial gas abundance profile {\em breaks}}
\label{breaks}
Some of the sample galaxies show a change in the slope of their radial $\log$(N/O)  and/or  $12+\log$(O/H)  profiles (as determined from at least one of the abundance diagnostics we adopted) that occurs at an average galactocentric radius of ($0.9\pm0.3$)$\times$r$_{25}$. Our criterion for the detection of such breaks was based on an improvement in the $\chi_\nu^2$ parameter in the double linear fit with respect to the single linear fit to the radial profile (see Sect.~\ref{fits} of this paper and sect.~5 of \citetalias{paperI} for further details).

A detailed analysis of the radial abundance profile breaks is beyond the scope of this paper, and in any case, further data will be needed in order to properly map the outer discs of the galaxies. However, we do find important to ensure that the change in slope, that usually implies a steepening in the inner disc profile and a flattening in the outer disc, does not affect the results presented in the previous sections. The number of galaxies with breaks is only  $\sim$20\% of the whole sample, roughly equally distributed between unbarred, weakly and strongly barred galaxies, with 4, 3 and 3 galaxies, respectively. In any case, in Fig.~\ref{NO_vs_M} we have plotted, with small symbols, the y-intercepts and the slopes of the $\log$(N/O) radial profiles corresponding to the inner disc in the bi-linear fit. The values obtained from the bi-linear fits and the corresponding ones from the single linear fits are connected with segments. As we can see the distributions do not change considerably. The linear fit to the $\alpha_{N/O}$ - M$_B$ relation remains the same within errors, with a slightly worse correlation coefficient (r=$-0.84$) for N/O abundances derived from HCM, and a slightly better coefficient for the relation derived from the $R$ calibration (r=$-0.86$). We have done the same exercise for the oxygen abundances determined with N2O2, as it is one of the strong-line methods for which the break in the metallicity profile is more clearly visible. The slopes and central abundances for the linear fit of the inner galaxy region is shown in the corresponding panel in Fig.~\ref{empi_vs_M} with small symbols. The same trend as for the single linear fit holds. Therefore, we conclude that the presence of radial breaks in the radial abundance profiles of the galaxies in our sample does not affect  the results presented in the previous sections.

%%%%%%%%%%%%%%%%%%%%%%%%%%%%%%%%%%%%%%%%%%%%%%%%%%%%%%%%%%%%%%%%%%%%%%%%%%%%%%%%%%%%%%%%%%%%%%%%%%%%%%%%%%%%%%
%%%%%%%%%%%%%%%%%%%%%%%%%%%%%%%%%%%%%%%%%%%%%%%%%%%%%%%%%%%%%%%%%%%%%%%%%%%%%%%%%%%%%%%%%%%%%%%%%%%%%%%%%%%%%%
%%%%%%%%%%%%%%%%%%%%%%%%%%%%%%%%%%%%%%%%%%%%%%%%%%%%%%%%%%%%%%%%%%%%%%%%%%%%%%%%%%%%%%%%%%%%%%%%%%%%%%%%%%%%%%
\subsection{Inclusion/exclusion of H\,{\small II} regions located within bars}
Some of the \hii\ regions whose fluxes have been compiled in this study are located within the bars of the host barred galaxies. If the \hii\ regions located within bars had different chemical properties than those located in the galactic discs, this could produce differences in radial gradients between barred and unbarred galaxies. 
%These differences would, in part or completely, be driven by the \hii\ regions located within bars. 

We have therefore repeated the same analysis as in the previous sections, but now considering the least-squares linear regression parameters derived when excluding from the fits all \hii\ regions located within bars. These regions are marked with a dark grey edge symbol in the plots of appendix~E in \citetalias{paperI}.  The histograms of y-intercepts and slopes (Figs.~\ref{histog_fits_HCM_NO}, \ref{histog_fits_HCM_OH} and \ref{histog_empi}) and corresponding AD-test results do not change, except for $\alpha_{N/O}$ for the $R$ calibration, yielding AD values below 5\% for slopes in dex~kpc\me\ and dex~r$_e$\me. The median values presented in Tables~\ref{tab:fits_NO} and \ref{tab:fits_OH} also remain unaffected  (within the errors). The uncertainties of the slopes and y-intercepts do increase in some cases, due to a smaller number of data points available for the fits. In particular, for two barred galaxies (NGC~1097 and NGC~5248) there are not enough regions outside their bars to obtain a reliable fit. The relations between slopes and y-intercepts with morphological  type, absolute magnitude and bar parameters (Figs.~\ref{empi_vs_M} to \ref{HCM_vs_bar_params}) do not change either.

The fact that the inclusion/exclusion of \hii\ regions located within bars does not affect the results implies that if the observed flattening is produced by the bar, its effect might be a large scale phenomenon that homogenises the gas chemical composition in the whole galactic disc, and  is not limited to the bar area.

%%%%%%%%%%%%%%%%%%%%%%%%%%%%%%%%%%%%%%%%%%%%%%%%%%%%%%%%%%%%%%%%%%%%%%%%%%%%%%%%%%%%%%%%%%%%%%%%%%%%%%%%%%%%%%
%%%%%%%%%%%%%%%%%%%%%%%%%%%%%%%%%%%%%%%%%%%%%%%%%%%%%%%%%%%%%%%%%%%%%%%%%%%%%%%%%%%%%%%%%%%%%%%%%%%%%%%%%%%%%%
%%%%%%%%%%%%%%%%%%%%%%%%%%%%%%%%%%%%%%%%%%%%%%%%%%%%%%%%%%%%%%%%%%%%%%%%%%%%%%%%%%%%%%%%%%%%%%%%%%%%%%%%%%%%%%
\section{Discussion}
\label{discusion}
Our analysis of the radial profiles of  $\log$(N/O) and $12+\log$(O/H) abundances in nearby spirals points to a luminosity-dependent difference between  barred and unbarred galaxies. High luminosity (M$_B\lesssim-19.5$) strongly barred and unbarred galaxies show similar abundance profile gradients. However, in the lower luminosity range the difference in abundance gradients between barred and unbarred galaxies increases with decreasing galaxy luminosity. The reason for this is a different dependence of the slopes with luminosity for the two galaxy types: high luminosity  strongly  barred galaxies show rather shallow $12+\log$(O/H) (with any diagnostic) and $\log$(N/O) radial abundance profiles (in dex~kpc$^{-1}$)  and this trend seems to hold for the whole luminosity range  ($-17\gtrsim$ M$_B\gtrsim-22$), in spite of the much lower number statistics  at low luminosity (M$_B\gtrsim-19.5$).   However, for unbarred galaxies the $\log$(N/O) and $12+\log$(O/H)  radial abundance profiles become steeper with decreasing luminosity (or mass). 

We do not detect differences in central $12+\log$(O/H) or $\log$(N/O)  abundances between barred and unbarred galaxies, as determined from the y-intercept of the linear fits, nor a dependence of central  abundances or slopes with current properties of the bars.

We start the discussion below by comparing our results with those  obtained by previous authors. We then continue with a review  of observational work and numerical models concerning the effect of stellar bars on the gas-phase of spirals and propose the possible scenarios that could explain our results. Next we review the current expectations for the abundance gradients from chemical evolution models and explore the influence of radial gas flows induced by bars and spirals.

%%%%%%%%%%%%%%%%%%%%%%%%%%%%%%%%%%%%%%%%%%%%%%%%%%%%%%%%%%%%%%%%%%%%%%%%%%%%%%%%%%%%%%%%%%%%%%%%%%%%%%%%%%%%%%
%%%%%%%%%%%%%%%%%%%%%%%%%%%%%%%%%%%%%%%%%%%%%%%%%%%%%%%%%%%%%%%%%%%%%%%%%%%%%%%%%%%%%%%%%%%%%%%%%%%%%%%%%%%%%%
%%%%%%%%%%%%%%%%%%%%%%%%%%%%%%%%%%%%%%%%%%%%%%%%%%%%%%%%%%%%%%%%%%%%%%%%%%%%%%%%%%%%%%%%%%%%%%%%%%%%%%%%%%%%%%
\subsection{Comparison with previous work}
Since \citet{pagel79} first suggested that barred spirals have flatter metallicity gradients than unbarred galaxies, many authors have tried to observationally confirm such a feature  in the radial  $12+\log$(O/H) profile.
%presumably produced by the large-scale gas flows induced by galactic bars.
Substantial work in this respect was done in the 1990's, with data samples of typically a few to up to $\sim40$ galaxies. The work by \citet{vila-costas92} was pioneering at studying \hii\ region properties in relation to global galaxy properties. These authors used a calibration of the R$_{23}$ parameter to obtain most of their metallicities, and found steep gradients for unbarred later-type galaxies, while all their barred galaxies (just five) had shallow gradients. Similar results were obtained by \citet{Zaritsky94}. They also used  R$_{23}$, with an average of three different calibrations \citep[but see][for some concerns on this calibration]{bresolin11_n4258} and reported a `{\em strong indication that barred galaxies do have flatter gradients than unbarred galaxies}' from a galaxy sample size alike the one in \citet{vila-costas92}, with 39 galaxies, including 6 barred spirals. However, \citet{Zaritsky94} reported that `{\em a few unbarred galaxies also have gradients significantly shallower than the average}'. This is in very good agreement with our findings. The galaxies included in   \citet{vila-costas92} and \citet{Zaritsky94}  are all nearby spirals covering a range in luminosity from M$_B\sim-17.5$ to $\sim-21.5$, comparable to ours. In fact, many of the galaxies in their samples are also included in our re-analysis. The similar luminosity range might be the reason for such an agreement.

Further work in the 1990's was done by \citet{MartinRoy92}, \citet{Martin94}, \citet{MartinRoy95}, \citet{Roy97} for individual galaxies or small galaxy samples.  \citet{Dutil_Roy} analysed a sample containing 16 (intermediate or strongly) barred galaxies. They combined \hii\ spectrophotometry and multi-slit data of spirals, with metallicity gradients obtained by their team or compiled from previous authors (and therefore obtained with different empirical calibrations). The metallicities for their observations were obtained from the [\oiii]/\hb\ and [\nii]/[\oiii] calibrations of \citet{Edmunds84}. 
\citet{Dutil_Roy} concluded that barred galaxies have shallower gradients than unbarred galaxies, but also noticed that early-type (presumably more massive) unbarred galaxies have  shallow gradients, similar to those  observed in strongly barred galaxies. This was also observed by \citet{MartinRoy92} in their analysis of the barred galaxy NGC~4303 from spectrophotometry of 79 \hii\ regions. As the metallicity gradient obtained by these authors is not flatter than that  of other unbarred spirals, the authors conclude that  the presence of the bar does not affect the relative distribution of elements in the disc of NGC 4303.
The authors explicitly compare NGC 4303 with NGC~628, NGC~2997 and the Milky Way (the latter  now known to be barred). All these galaxies have M$_B<-19.8$ and therefore belong to the brighter part of the luminosity range. In this luminosity range we do not see  differences between  barred and unbarred galaxies either. A novelty in \citet{Martin94} and \citet{Dutil_Roy}  is that they did not use  the RC3 morphological classification of bars, but rather specific measurements of the bar ellipticities.
%From this, they report a dependence among the oxygen gradient and the bar ellipticity and length. We disagree in this point, as we do not observe such relation.

After the 1990's the topic of the bar effects on the distribution of metals in spirals did not receive much attention until the past decade, with the advent of the Integral Field Spectroscopy (IFS) instruments. Their associated legacy surveys permitted to increase considerably the galaxy samples sizes as well as the number of \hii\ regions observed in a single galaxy. The CALIFA survey \citep{Walcher2014} has been especially  prolific in  this respect.  The first analysis of ionized gas metallicity gradients in spirals that emerged from this project, published by \citet{sanchez14}, included 193 galaxies. These authors did not find statistically significant differences in the slope of barred and unbarred galaxies. 
%Furthermore,   the authors concluded that disc galaxies have a universal gradient in oxygen abundance ($-0.1$~dex~r$_e$\me) when normalised to $r_e$ \citepalias[but see sect.~5.3 in][]{paperI}. 
The same conclusion was reached by \citet{Sanchez-Menguiano2016}, who measured oxygen abundances in a sub-sample of 122 CALIFA face-on galaxies. In this case the radial abundance profiles were obtained on a spaxel-by-spaxel basis (rather than being based on the analysis of discrete \hii\ regions or star-forming complexes). Their fig.~1 shows that all galaxies in their
sample are brighter than M$_B \approx -19$. We have found differences between barred and unbarred systems  only for lower luminosity galaxies (M$_B\gtrsim-19.5$). The lack of low luminosity galaxies in the CALIFA sample can be the reason why they find no bar effect on the gradients. It is also worth mentioning that both \citet{sanchez14} and \citet{Sanchez-Menguiano2016}\footnote{Also \citet{sanchez2012} in a previous work with $\sim38$ galaxies from the PINGS survey.} used the O3N2 parameter as calibrated by \citet{pp04} and \citet{marino13}, respectively. We saw in Fig.~\ref{empi_vs_M} that the differences between barred and unbarred galaxies were less clear with these calibrations, presumably due to 
%\col{the fact that the O3N2 diagnostic yields radial metallicity profiles of the galaxies with the steepest gradients that are shallower compared to other diagnostics 
the fact that the O3N2 diagnostic, when compared to  other diagnostics, yields shallower radial metallicity profiles for the galaxies with the steepest gradients (see \citetalias{paperI} for a detailed analysis).  \citet{Perez-Montero2016} and \citet{Zinchenko2019} also analysed ionized gas metallicity gradients for different sub-samples of CALIFA galaxies, for \hii\ regions and individual spaxels, respectively. Both studies found no significant differences in the radial oxygen abundance gradient between barred and unbarred galaxies. They used  strong-line methods other than O3N2 to derive their abundances \citep[HCM and $R$ calibration of][]{PilyuginGrebel2016}.  We therefore believe that, although the strong-line method employed can affect the details,  the luminosity distribution of the CALIFA galaxies might be the dominant reason why these works did not detect differences in the metallicity gradients of barred vs. unbarred galaxies. Our results do agree with \citet{Zinchenko2019} in that high mass (barred or unbarred) galaxies  exhibit a rather flat oxygen abundance gradient (in units of dex~kpc$^{-1}$).

\citet{Kaplan2016} also reached the conclusion that the presence of bars does not affect galactic oxygen abundance gradients. Their work is also based on IFS data, with improved spatial resolution with respect to CALIFA \citep{sanchez14}. Their spaxel-to-spaxel abundance estimates are carefully done in order to exclude from their analysis regions dominated by Seyfert, LINER or diffuse ionized gas emission, and use seven different strong-line oxygen abundance indicators to obtain the metallicity gradients. They find that `{\em isolated barred and unbarred spiral galaxies exhibit similarly shallow 12 + $\log$(O/H) profiles}'. Using the N2O2  diagnostic developed by \citet{kewley02} they obtain average gradients of $-0.44\pm0.13$ and $-0.33\pm0.01$ dex~r$_{25}^{-1}$, respectively, for isolated unbarred and barred spirals, that agree within errors. These conclusions are based on a small sample of just eight galaxies, with the  statistics above based on only four unbarred galaxies (NGC~628, NGC~1068, NGC~3938 and NGC~4254) and two isolated barred galaxies (NGC~2903, NGC~5713). The stellar masses of these galaxies are all above $2\times10^{10}$~M$_{\odot}$, and their absolute $B$-band magnitudes $<-20$. In such a range of luminosities we do not see  differences in abundance gradients between barred and unbarred galaxies either. In fact, at least four of the galaxies studied by \citet{Kaplan2016} are included in our sample, and our determination of their $\log$(O/H) radial  gradients agrees rather well with their work, in spite of our lower spatial coverage of the discs. Our result is therefore not in contradiction with \citet{Kaplan2016}, but rather in agreement for the specific range of stellar mass (or total luminosity) of their sample galaxies.

\begin{figure}
\includegraphics[width=0.5\textwidth]{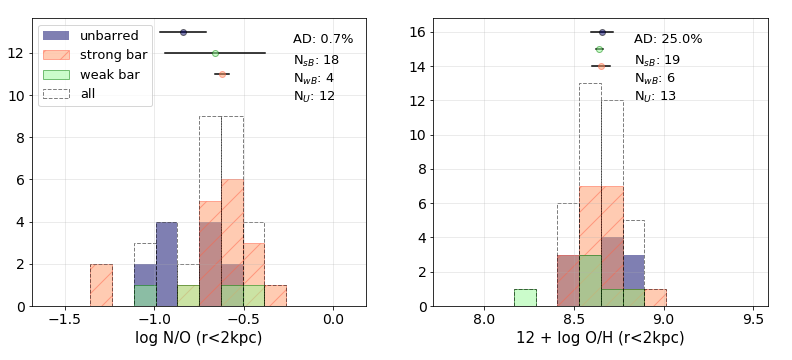}
\caption{\label{inner_abund} Distribution of the average $\log$(N/O) ({\em left}) and $12+\log$(O/H)  ({\em right}) within the inner region of the galaxies of the sample. The average abundances have been obtained by calculation of the mean abundances of the \hii\ regions with galactocentric radius $<$~2~kpc. Abundances here were obtained from the HCM method.}
\end{figure}

Comparative works on the $\log$(N/O) abundances in barred and unbarred galaxies are far more scarce. To our knowledge, only \citet{Perez-Montero2016} compared the $\log$(N/O) radial gradients in barred and unbarred galaxies from the CALIFA survey, and as for the metallicity gradients, they found no difference. These authors do find a trend for barred galaxies to show larger central  $\log$(N/O), but differences are within errors. A significant difference in $\log$(N/O) between barred and unbarred galaxies was instead found by \citet{bulbos}. The ionized gas abundances were derived in this case from SDSS spectroscopy of the inner 3\arcsec\ of spirals (or  inner 2-4~kpc, taking sample galaxy distances into account). The $\log$(N/O) was found to be enhanced in the centre of barred galaxies with respect to unbarred ones, after removing AGNs. The observed differences between barred and unbarred galaxies were larger in the less massive galaxies (M$_\star \leq 10^{10.8}$~M$_\odot$). The y-intercepts of our fits to the  $\log$(N/O) radial profiles show no significant difference (Fig.~\ref{histog_fits_HCM_NO}). However, in Fig.~\ref{NO_vs_M}, if we ignore Magellanic-type barred galaxies (marked with  grey edge symbols and showing the lowest  $\log$(N/O) values), there seems to be a trend for lower luminosity barred galaxies to have larger central  $\log$(N/O). In order to better compare our present results with those in \citet{bulbos}, we have created Fig.~\ref{inner_abund}. It shows histograms of the {\em central} $\log$(N/O) and $12+\log$(O/H) for the galaxies of the sample, derived from HCM as in \citet{bulbos}.  These central values were obtained by averaging  abundances of the \hii\ regions located in the inner parts of the galaxies (galactocentric radius $<$2~kpc, to emulate the SDSS results). Figure~\ref{inner_abund}  shows no difference in central $12+\log$(O/H) between barred and unbarred galaxies ({\em right}), but barred galaxies have  $\sim0.2$~dex larger central $\log$(N/O) compared to  unbarred galaxies ({\em left}, with AD $P$-value $\sim$0.7\%). In spite of the lower number statistics here, this is compatible with the result in \citet{bulbos}. The reason why a larger central average $\log$(N/O) is measured in barred galaxies, that is not detected in $12+\log$(O/H), might be due to a combination of factors:  the barred/unbarred difference in radial gradients  is better seen in $\log$(N/O) than in $12+\log$(O/H), there is a subtle trend for a larger y-intercept in the $\log$(N/O) profiles of barred galaxies and $\log$(N/O) gradients are shallower in barred galaxies. 

In summary, the fact that the O/H and N/O ratios differences observed between  barred and unbarred galaxies depend on galaxy luminosity (or mass) can explain the controversy characterizing previous studies.  The comparison with previous works shows that the range of galaxy luminosities sampled by different investigators might be the key element to explain the origin of the discrepant results, although we can not discard contributions arising from the use of different methodologies and/or abundance diagnostics.

%% Este grafico excluye regiones con 'A', por BPT o por estar en zona central de galaxia con AGN

%%%%%%%%%%%%%%%%%%%%%%%%%%%%%%%%%%%%%%%%%%%%%%%%%%%%%%%%%%%%%%%%%%%%%%%%%%%%%%%%%%%%%%%%%%%%%%%%%%%%%%%%%%%%%%
%%%%%%%%%%%%%%%%%%%%%%%%%%%%%%%%%%%%%%%%%%%%%%%%%%%%%%%%%%%%%%%%%%%%%%%%%%%%%%%%%%%%%%%%%%%%%%%%%%%%%%%%%%%%%%
%%%%%%%%%%%%%%%%%%%%%%%%%%%%%%%%%%%%%%%%%%%%%%%%%%%%%%%%%%%%%%%%%%%%%%%%%%%%%%%%%%%%%%%%%%%%%%%%%%%%%%%%%%%%%%
\subsection{Why is  the bar  expected to affect the gradients?}
\label{scenarios}
The results obtained in this work show that barred galaxies have shallow radial $\log$(O/H) and $\log$(N/O) abundance profiles at high luminosity and, although with lower number statistics, this trend seems to hold also in the lower luminosity range. The average slope for $\log$(O/H)  in the range covered by this galaxy sample ($-22<M_B<-17$), is in  the range [$-0.022$,$-0.004$] in dex~kpc$^{-1}$, with a typical dispersion of $\sim$0.01 dex~kpc$^{-1}$, where the range covers the values obtained with the different strong line methods. For $\log$(N/O), the average radial profile slope for barred galaxies is also considerably shallow, with a median value in the range [$-0.04$,$-0.02$] in dex~kpc$^{-1}$ depending on the method.

A shallow radial abundance profile in barred spirals is expected, at least qualitatively, according to simulations, due to the non-axisymmetric gravitational potential. The bar induces an angular momentum redistribution of the gas and stars in galactic discs \citep[e.g.][]{SW93,athanassoula03,friedli_benz}. The gas is dissipative and the angular momentum loss induced by the bar gives way to gas flows within the disc in different directions depending on the location: gas located within the corotation resonance radius (or approximately at the bar radius), with lower metallicity than the inner galaxy is driven inwards, towards the galaxy centre, while gas located beyond corotation is driven outwards, towards the outer part of the galactic disc, where gas is expected to have the lowest metallicity \citep[e.g.][]{Simkin,kormendy,SW93,friedli_kenni}.
Bar-induced gas inflow rates are in the range $\sim$0.1-4~M$_\odot$~yr$^{-1}$ according to
%Antes 0.1-1, pero Quillen mide 4 Msun/yr
observations \citep[e.g.][]{regan97,sakamoto,Quillen1995,SormaniBarnes19} and simulations \citep[e.g.][]{ReganTeuben,Maciejewski2002,cole}, but gas flows seem to have a smaller impact beyond the bar radius \citep{Kubryk2013}. Predicted gas inflow rates are highly dependent on numerical simulation schemes and parameters
\citep[e.g.][]{KimStone2012}; %Debe ser Kim et al. 2012 de Sellwood, p.31
in particular, they are  strongly dependent on the bar ellipticity (or bar axis ratio). \citet{ReganTeuben} find very little net gas inflow for bar axis ratios below 1.5 (or bar ellipticities below $\sim0.33$).
%% Puede ser util Section 6.1 de Friedli & Benz 1995, sobre relacion gradiente con b/a de la barra.
%% Martin & Roy 1995, encuentran una dependencia de pendiente con b/a de la barra

The real scenario may be far more complex than summarised here, as the bar strength changes with time \citep[][]{athanassoula13,cole}. Strong gas inflows are expected soon after the bar forms, and will decrease with time as the bar evolves, together with  the induced star formation along the bar and especially  in the innermost region of the galaxy, in either a nuclear disc or ring or a central starburst \citep[e.g.][]{cole,SandersTubbs1980,kormendy}. In the inner kiloparsec, the enhanced star formation induced by the inflows would lead to an enhanced production of metals and possibly also to enriched gas outflows driven by starbursts  \citep[e.g.][]{Krumholz2017}. The current radial distribution of metals will then be the result of the cumulative effect of the (local) star formation history and the gas inflows and outflows. These will  redistribute gas with a presumably initial radially decreasing metal abundance, and will therefore re-shape the initial radial abundance profile. According to simulations, the bar  homogenises the gas very efficiently, in just a few bar rotations, corresponding to time-scales on the order of a few hundred Myr \citep[e.g.][]{friedli_benz1995,isa2006}.
% $\sim200-300$~Myr the. Look for references.
The flattening of the abundance gradients in barred galaxies is therefore at least qualitatively explained by the mixing induced by the bars.

However, if the bar is the agent that flattens radial abundance gradients, why do massive unbarred galaxies show gradients as flat as those seen in strongly barred galaxies? What is the agent or process that flattens the profiles for high-luminosity unbarred galaxies? Are there different dominant flattening agents depending on the galaxy mass? If there were no differences in the radial abundance gradient between barred and unbarred galaxies at lower luminosity, a tentative conclusion would had been that the present-day stellar bars observed in spirals are not a dominant agent in flattening their abundance gradients. However, if we look at spirals in the lower mass/luminosity range we do see clear differences in gradients between  barred and unbarred galaxies. The dependence of the barred/unbarred gradient difference on the galaxy luminosity (or mass) opens interesting questions and suggests two (extreme) basic scenarios or interpretations:
\begin{enumerate}
\item Strong bars flatten radial abundance profiles in spirals, no matter what their mass is. In unbarred galaxies the abundance gradient (in dex~kpc\me) depends on the galaxy mass, being as flat as in barred galaxies for the most massive spirals. In this scenario,  an evolutionary process or an agent(s) different from a bar might be in action in massive unbarred galaxies, so that it flattens their radial abundance profiles.
 
\item Galaxies, as a consequence of their evolution, show a radial abundance profile that depends on their luminosity (mass), being steeper for the less luminous galaxies (next section). There is no need to invoke the flattening effect of bars in massive barred galaxies, as unbarred massive galaxies show the same flat gradient slopes. If any additional agent  contributes to the flattening in massive galaxies this might be equally efficient whether galaxies have a bar or not.
% There might be another agent(s) or evolutionary process (different from the bar) that makes radial abundance profiles shallow
% in massive spirals, either they have a bar or not.
  However, strong bars in low-mass galaxies might be efficient at mixing and redistributing metals, as  unbarred spirals of the same mass show considerably steeper radial abundance profiles.
\end{enumerate}

In the next subsections we review the relation between abundance gradient (in units of dex~kpc$^{-1}$) and galaxy luminosity in the context of chemical evolution models. We also explore the effects of gas flows in altering abundance gradients according to models, and how radial gas flows induced by bars or spiral patterns could work differently, depending on galaxy mass/luminosity.

%%%%%%%%%%%%%%%%%%%%%%%%%%%%%%%%%%%%%%%%%%%%%%%%%%%%%%%%%%%%%%%%%%%%%%%%%%%%%%%%%%%%%%%%%%%%%%%%%%%%%%%%%%%%%%
%%%%%%%%%%%%%%%%%%%%%%%%%%%%%%%%%%%%%%%%%%%%%%%%%%%%%%%%%%%%%%%%%%%%%%%%%%%%%%%%%%%%%%%%%%%%%%%%%%%%%%%%%%%%%%
%%%%%%%%%%%%%%%%%%%%%%%%%%%%%%%%%%%%%%%%%%%%%%%%%%%%%%%%%%%%%%%%%%%%%%%%%%%%%%%%%%%%%%%%%%%%%%%%%%%%%%%%%%%%%%
\subsection{The abundance gradient -  M$_{B}$ relation}
Figs.~\ref{empi_vs_M}  to \ref{NO_vs_M_otros} show that in unbarred galaxies  both the $\log$(O/H) and the  $\log$(N/O) abundance gradients (in units of dex~kpc$^{-1}$) steepen when decreasing galaxy luminosity (Sect.~\ref{fits_vs_MB}). This result is compatible with a common abundance gradient in dex~r$_e^{-1}$ for unbarred galaxies, given the positive relation between galaxy luminosity and size \citep{vandenbergh}.
 
%%%%%%%%%%%%%%%%%%%%%%%%%%%%%%%%%%%%%%%%%%%%%%%%%%%%%%%%%%%%%%%%%%%%%%%%%%%%%%%%%%%%%%%%%%%%%%%%%%%%%%%%%%%%%%
%%%%%%%%%%%%%%%%%%%%%%%%%%%%%%%%%%%%%%%%%%%%%%%%%%%%%%%%%%%%%%%%%%%%%%%%%%%%%%%%%%%%%%%%%%%%%%%%%%%%%%%%%%%%%%
%%%%%%%%%%%%%%%%%%%%%%%%%%%%%%%%%%%%%%%%%%%%%%%%%%%%%%%%%%%%%%%%%%%%%%%%%%%%%%%%%%%%%%%%%%%%%%%%%%%%%%%%%%%%%%
\subsubsection{O/H abundance gradient}
\label{alpha_OH_MB}
A relation between the oxygen abundance gradient and the galaxy stellar mass or luminosity has been found by a number of authors \citep[e.g.][]{vila-costas92,Garnett97,PB00,Ho2015,Bresolin2019} in local galaxies, with low-luminosity galaxies having on average a steeper $\log$(O/H) radial gradient in dex~kpc$^{-1}$.

\citet{Carton2018} also find a dependence with galaxy size, in the form of a  larger spread in observed metallicity gradients for small galaxies at $0.08 <z < 0.84$, while large galaxies show less scatter. A larger scatter for lower values of log~r$_{25}$ is also found  by \citet{pilyugin2014} for nearby galaxies.  The larger spread in gradients at low luminosity is present at absolute magnitudes M$_B\gtrsim-19$ -- see fig.~11 of \citet{Bresolin2019} or fig.~12 in \citet{Ho2015} with a tendency for steeper gradients for decreasing luminosity.

A steeper radial metallicity gradient in dex~kpc$^{-1}$ for lower-luminosity (or lower-mass) galaxies is predicted by analytical chemical evolution models \citep{PB00,MollaDiaz05,Molla19},  where it appears as a consequence of assuming a star formation law that depends on surface gas density (Schmidt law) that  decreases with radius. This produces an {\em inside-out} growth of the discs. In  chemodynamical and cosmological simulations the inside-out disc growth appears as a natural outcome resulting in a steepening of the metallicity profiles for smaller or less massive galaxies \citep{Gibson2013,Pilkington2012,Tissera2016,Tissera2019,Few2012}. A consequence of the inside-out star-forming scheme is a flattening of the radial abundance profile with time: the profile is steep early on when stars and metals are forming only in the inner disc, and flattens as the star formation moves progressively outwards \citep[but see][]{Chiappini2001}. In addition, the  inside-out mechanism implies an evolution timescale dependence with galaxy mass: massive galaxies form their stars and evolve first, while low-mass galaxies evolve more slowly and are therefore expected to show steeper present-day radial abundance profiles than massive galaxies.

%%%%%%%%%%%%%%%%%%%%%%%%%%%%%%%%%%%%%%%%%%%%%%%%%%%%%%%%%%%%%%%%%%%%%%%%%%%%%%%%%%%%%%%%%%%%%%%%%%%%%%%%%%%%%%
%%%%%%%%%%%%%%%%%%%%%%%%%%%%%%%%%%%%%%%%%%%%%%%%%%%%%%%%%%%%%%%%%%%%%%%%%%%%%%%%%%%%%%%%%%%%%%%%%%%%%%%%%%%%%%
%%%%%%%%%%%%%%%%%%%%%%%%%%%%%%%%%%%%%%%%%%%%%%%%%%%%%%%%%%%%%%%%%%%%%%%%%%%%%%%%%%%%%%%%%%%%%%%%%%%%%%%%%%%%%%
\subsubsection{N/O abundance gradient}
Observational work on the $\log$(N/O) radial gradient in galaxies is far more scarce  than for O/H. In particular, to our knowledge, only \citet{considere} previously inspected the $\log$(N/O) radial gradient (in dex~kpc$^{-1}$) as a function of luminosity for a sample of barred starbursts.

\begin{figure}
\includegraphics[width=0.5\textwidth]{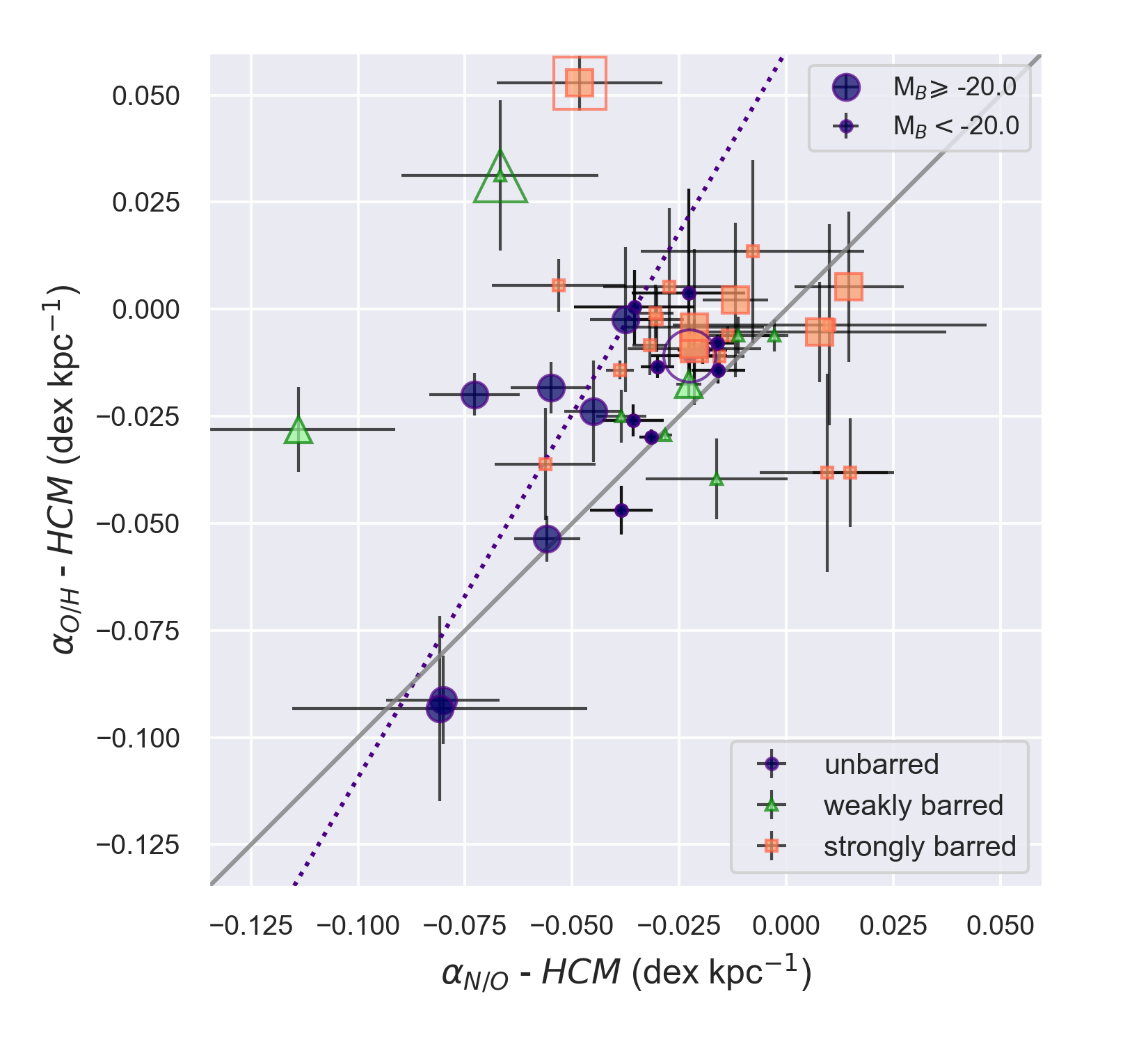}
\caption{Slope of the  $12+\log$(O/H) radial abundance profile (from HCM) as a function of the  slope of the $\log$(N/O) radial profile for the galaxies of the sample. Circles, triangles and  squares represent strongly barred, weakly barred and unbarred galaxies, respectively. Galaxies with low luminosity (M$_B \geqslant-20$) are  marked with big symbols. The dotted-straight line is a linear fit to the unbarred low-luminosity galaxies, and the  continuous grey line marks the 1:1 relation.  Big empty symbols mark galaxies for which the slope is defined with only five to seven data points, and are therefore less reliable. \label{alphaNO_vs_alphaOH}}
\end{figure}

The evolution of $\log$(N/O)  has been studied observationally and theoretically for individual \hii\ regions and for whole star-forming galaxies, mostly as an attempt to understand the evolution in the $\log$(N/O) - $\log$(O/H) plane as a function of stellar yields and  star formation efficiency, among others \citep[e.g.][]{MollaVilchez2006,Vincenzo2016,pmc09}. However, observational work on the dependence of radial gradients in $\log$(N/O)  with global galaxy properties is more scarce \citep[e.g.][]{Perez-Montero2016,Belfiore2017}.
\citet{Belfiore2017} used N2O2 to measure $\alpha_{N/O}$ for 550 nearby galaxies from the SDSS IV MaNGA survey and report a steepening of the profiles as a function of galaxy mass, for slopes in dex~r$_{e}$\me. The steepening represents a rather small change, 0.02~dex~r$_{e}$\me, over the whole mass range (from $10^9$-$10^{11}$M$_\odot$), being much smaller than the average scatter of the slopes, of 0.09~dex~r$_{e}$\me. We do not detect that steepening in mass. Our Fig.~\ref{NO_vs_M} (bottom-right) shows a rather constant slope across the whole luminosity range (that roughly corresponds to the range in mass covered by \citet{Belfiore2017}). However, our median slope for the whole sample,  $-0.16$ and $-0.15$~dex~r$_{e}$\me\  for HCM and the $R$ method, respectively, is very close to the value reported by \citet{Belfiore2017}, in spite of the fact that a different strong-line method is used. Our results do agree  also with \citet{Belfiore2017} in that the   metallicity  gradients are slightly shallower than the $\log$(N/O) radial gradients. This can be seen in Fig.~\ref{alphaNO_vs_alphaOH}, which correlates the O/H and the N/O gradient slopes measured in our sample. The same plot shows also a tendency for unbarred galaxies to have correlated metallicity and $\log$(N/O) gradients ($\rho$=0.68), that improves ($\rho$=0.82) when we only consider low-luminosity galaxies (M$_B \geqslant -20$). The linear fit constrained to unbarred spirals with M$_B \geqslant -20$ is shown with a straight dotted blue line:
\begin{equation}
\alpha_{O/H} = (0.06\pm0.05) +  (1.7\pm0.8) ~ \alpha_{N/O}
\end{equation}
which has a correlation coefficient of 0.8.

The analysis of the evolution of $\alpha_{N/O}$ in galaxies from cosmological hydrodynamical simulations from \citet{Vincenzo2018}, predicts a flattening of  the gradient (in dex~kpc\me) with time, that under the inside-out scenario agrees with our results, as it  implies a steeper profile (and a slower evolution) for low-mass galaxies. To our knowledge there are no further predictions from analytical or chemodynamical evolution models on the relation between $\alpha_{N/O}$ and galaxy mass or luminosity.
%, but given the observed correlation
% between  $\log$(N/O) - $\log$(O/H)  at high metallicity

%%%%%%%%%%%%%%%%%%%%%%%%%%%%%%%%%%%%%%%%%%%%%%%%%%%%%%%%%%%%%%%%%%%%%%%%%%%%%%%%%%%%%%%%%%%%%%%%%%%%%%%%%%%%%%
%%%%%%%%%%%%%%%%%%%%%%%%%%%%%%%%%%%%%%%%%%%%%%%%%%%%%%%%%%%%%%%%%%%%%%%%%%%%%%%%%%%%%%%%%%%%%%%%%%%%%%%%%%%%%%
%%%%%%%%%%%%%%%%%%%%%%%%%%%%%%%%%%%%%%%%%%%%%%%%%%%%%%%%%%%%%%%%%%%%%%%%%%%%%%%%%%%%%%%%%%%%%%%%%%%%%%%%%%%%%%
\subsection{Effect of radial flows on abundance profiles}
The \citet{BP99} and \citet{MollaDiaz05} chemical evolution models mentioned in Sect.~\ref{alpha_OH_MB} include neither radial flows nor outflows. They were created to reproduce Milky Way observables. The  \citet{BP99} model was afterwards extended to other discs  by adopting scaling relations from the framework of cold dark matter semi-analytic models of galaxy formation \citep{Mo1998}. Their predictions  were in good agreement with observational data (for unbarred galaxies) available by then. However, \citet{Bresolin2019} recently compared their model predictions  with a larger sample of galaxies and found  that, although observations and model qualitatively agree, the model predicts radial metallicity profiles that are systematically steeper than observed, which  makes the author conclude that it may be due to the absence of important processes such as gas radial flows and outflows in this kind of models. \citet{Kaplan2016} also found discrepancies, in the same direction as \citet{Bresolin2019}, when comparing  metallicity gradients on a smaller sample of 8 local galaxies and two galaxies at redshifts 1.5 and 2, with the predictions of \citet{PB00}.

In fact, radial flows can be one of the most important processes that shape the abundance gradient. These have been included in analytical chemical evolution models of the Galaxy \citep[e.g.][]{SchonrichBinney2009, SpitoniMatteucci2011,GrisoniSpitoniMatteucci2018} and naturally appear in simulations including strong feedback \citep{Gibson2013}. Although there is no need to include radial flows to reproduce the present-day Milky Way profile under the inside-out scheme (a variable star formation efficiency and threshold in gas density are enough), these are required to simultaneously reproduce both the metallicity gradient and the radial gas density profile in the inner disc \citep{PortinariChiosi1999}.

%%%%%%%%%%%%%%%%%%%%%%%%%%%%%%%%%%%%%%%%%%%%%%%%%%%%%%%%%%%%%%%%%%%%%%%%%%%%%%%%%%%%%%%%%%%%%%%%%%%%%%%%%%%%%%
%%%%%%%%%%%%%%%%%%%%%%%%%%%%%%%%%%%%%%%%%%%%%%%%%%%%%%%%%%%%%%%%%%%%%%%%%%%%%%%%%%%%%%%%%%%%%%%%%%%%%%%%%%%%%%
%%%%%%%%%%%%%%%%%%%%%%%%%%%%%%%%%%%%%%%%%%%%%%%%%%%%%%%%%%%%%%%%%%%%%%%%%%%%%%%%%%%%%%%%%%%%%%%%%%%%%%%%%%%%%%
\subsubsection{Bar-induced flows}
Bar-induced gas flows are more difficult to simulate. \citet{Cavichia2014} modified the \citet{MollaDiaz05} model to incorporate radial gas flows in order to mimic the effect of the Galactic bar. The model predicts a flattening of the metallicity profile, but the difference is very small with respect to the model with no bar, in part due to the rapid early evolution of the disc, that establishes a rather flat profile early on, even before the bar formation (assumed to occur 3~Gyr ago). However, the model with a bar better reproduces  the surface gas density and the star formation rate profiles. A similar conclusion is drawn by \citet{Kubryk2013} from N-body + smoothed particle hydrodynamics. They investigate the effect of the bar-induced radial migration of gas and stars on the galaxy chemical evolution. Despite the important amount of radial migration occurring in their model, its impact on the chemical properties is limited. Their simulated Milky-Way mass disc suffers a flattening in the oxygen abundance profile of only 0.01~dex~kpc\me\ (from an initial $-0.03$ up to $-0.02$~dex~kpc\me). However, the authors identify a  number of factors that control the impact of bar-induced radial migration: (i) the strength of the bar gravitational potential (a stronger bar favours larger effects), (ii) the duration of the radial migration (the longer the bar acts, the larger the effects), and (iii) the steepness of the initial abundance profile. With regards to the latter the authors do the experiment of setting in an initial $-0.08$~dex~kpc\me\ oxygen gradient, and the mixing induced by the same simulated bar decreases the gradient down to $-0.04$~dex~kpc\me, i.e. it produces a much  flatter gradient compared to  the initial simulation. If we assume that the flattening observed in barred galaxies is due to the bar-induced mixing, the above results point to a larger effect on the less massive/luminous galaxies, where the initial metallicity profile is expected to be steeper (Figs.~\ref{empi_vs_M}, \ref{NO_vs_M} and \ref{NO_vs_M_otros}).
A larger effect of the bar in low-mass barred galaxies with respect to high-mass galaxies is found by \citet{Martel2013} and \citet{Martel2018} from a  smoothed particle hydrodynamics algorithm specially designed to simulate galactic chemodynamical evolution. However, these authors do not analyse metallicity gradients but just central metallicity values. To our knowledge there are no  further model predictions on the  bar-induced metallicity profile flattening in lower mass-galaxies. Galaxy numerical simulations are also usually done to reproduce Milky Way-mass galaxies, and the total mass is usually kept constant. However, the bar-induced gas inflow rate has been measured in simulations as a function of the bar strength and quadrupole moment by \citet{athanassoula} and \citet{ReganTeuben}. For a fixed total galaxy mass and a given bar ellipticity, the gas inflow rate towards the inner kiloparsec increases for bars with a larger quadrupole moment. In the mentioned conditions, a larger quadrupole moment implies a larger  mass for the bar relative to the disc. In addition, a larger inflow rate in a lower mass galaxy would dilute the metals more rapidly. A higher efficiency in the bar-induced mixing in a low mass disc  is, at least qualitatively, compatible with simulations. 

%%%%%%%%%%%%%%%%%%%%%%%%%%%%%%%%%%%%%%%%%%%%%%%%%%%%%%%%%%%%%%%%%%%%%%%%%%%%%%%%%%%%%%%%%%%%%%%%%%%%%%%%%%%%%%
%%%%%%%%%%%%%%%%%%%%%%%%%%%%%%%%%%%%%%%%%%%%%%%%%%%%%%%%%%%%%%%%%%%%%%%%%%%%%%%%%%%%%%%%%%%%%%%%%%%%%%%%%%%%%%
%%%%%%%%%%%%%%%%%%%%%%%%%%%%%%%%%%%%%%%%%%%%%%%%%%%%%%%%%%%%%%%%%%%%%%%%%%%%%%%%%%%%%%%%%%%%%%%%%%%%%%%%%%%%%%
\subsubsection{Spiral pattern-induced mixing}
Apart from galactic large-scale stellar bars, the spiral pattern can also produce gas angular momentum transport  mainly via shocks, and therefore generate  large-scale gas flows within galactic discs  \citep[][]{Lubow1986,SellwoodBinney2002,Hopkins2011,KimKim2014}. The gas would flow inwards or outwards, depending on  the location. Outside the corotation radius resonance, the shocks at the location of the spiral arm provide a positive torque and drive the gas towards larger radii, whereas inside the corotation radius,  the gas loses angular momentum in the shocks and is driven inwards \citep[][]{Shu1992}.
%depending on the sign of $\Omega-\Omega_p$, where $\Omega$ and $\Omega_p$ are the
%circular frequency of the gas and the spiral pattern speed, respectively. Outside the corotation radius resonance,
%where  $\Omega < \Omega_p$, the shocks at the spiral provide a positive torque and drive the gas towards larger radius,
%whereas inside the corotation radius, where  $\Omega > \Omega_p$, the gas looses angular momentum
%in the shocks and is driven inwards \citep[][]{Shu1992}. %Check others, also Binney & TRemaine
%\citet{Lubow1986} has estimated spiral induced gas accretion rate towards the Milky Way centre of 0.2-0.4~M$_\odot$~yr\me.
\citet{KimKim2014} made hydrodynamical simulations to investigate the gas response to a spiral potential in a disc galaxy
and, in particular, to quantify the mass inflow rate induced by the spiral.
%The spiral pattern is characterized by two parameters:
%the angular frequency or pattern speed ($\Omega_p$) and the spiral strength, $\mathcal{F}$.
They estimate inflow rates in the range $\sim0.5-3$~M$_\odot$~yr\me, with the larger rates corresponding to the slower and stronger spiral patterns. For a fixed amplitude in the spiral gravitational potential, the spiral strength is larger for spirals with a larger number of arms, but has a stronger dependence with the arms pitch angle. Then, in the absence of a bar, and for galaxies with identical number of arms and spiral gravitational potential, those   with lower pitch angles (or tighter arms) are expected to exert a stronger force on  the gas in the direction perpendicular to the arms,
%relative to the radial force from the background axisymmetric potential,
compared to  spiral galaxies presenting looser arms.  Late-type spirals tend to have larger pitch angles than early-type galaxies \citep[e.g.][]{Kennicutt1981,Treuthardt2012,GarciaGomez1993,simon2019}. In addition, observations reveal a relation between the number and class of spiral arms and their pitch angles, in the sense that many-armed and flocculent spiral arms are looser than two-armed grand design structures \citep{GarciaGomez1993,Hart2017}.
%Grand design and multiple-arm spirals are very likely to be present in early-type galaxies, whereas
Flocculent galaxies are mostly late-type, low-luminosity systems \citep{Elmegreen1982,Ann2013}.
%\citet{Elmegreen} find a deficit of grand design galaxies in Sd and Sm types, and \citet{Ann} find that flocculent galaxies are mostly late types 
The aforementioned results imply that in unbarred grand-design spirals, important gas flows may be created, with inflow rates similar to those produced by strong bars, while in later-type unbarred galaxies, where multiple arms and flocculent spirals are more frequent, the spiral pattern strength is lower, and the impact of spiral-induced gas flows is expected to  be considerably smaller \citep{KimKim2014}.

\begin{figure*}
\includegraphics[width=0.80\textwidth]{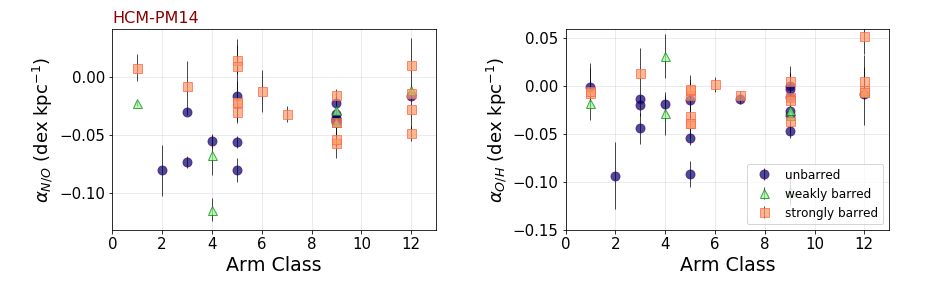}
\caption{{\em Left:} Slope of the $\log$(N/O)  radial abundance profile (from HCM) as a function of the  arm class according to \citet{Elmegreen1987}.  {\em Right:} Same but for the slope of the  $12+\log$(O/H) radial abundance profile.\label{AC}}
\end{figure*}

\begin{figure*}
\includegraphics[width=0.80\textwidth]{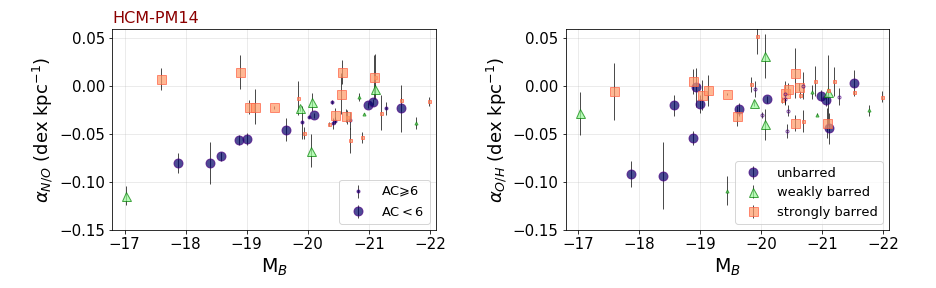}
\caption{{\em Left:} Slope of the $\log$(N/O)  radial abundance profile (from HCM) as a function of the $B$-band absolute magnitude (M$_B$). {\em Right:} Same but for the slope of the $12+\log$(O/H) radial abundance profile. Small and big symbols show galaxies with arm class larger or smaller than 6, respectively.
 \label{AC2}}
\end{figure*}

Our results support this hypothesis as we show below. The unbarred galaxies in our sample with M$_{B}\gtrsim -19.5$ are  NGC~300, NGC~1068, NGC~2403, NGC~2541, NGC~598 and NGC~7793. According to the arm classification (hereinafter AC) by \citet{Elmegreen1987}, they have classes in the range 1-5, and can be therefore referred to as flocculent (class 1-3) or multiple arm (classes 4-9) \citep{Elmegreen2011}  systems. Fig.~\ref{AC}  shows the slope of the radial  $\log$(N/O) and 12$+\log$(O/H) profiles (as derived from HCM) as a function of the AC, in the left and right panels, respectively. The plot shows a trend for galaxies with AC above 6 to have rather shallow profiles, while the dispersion increases considerably for those with lower arm classes. The latter galaxies cover the whole range of observed slopes. In Fig.~\ref{AC2}  the slopes are plotted as a function of M$_B$ and we have indicated with big symbols of different colour the galaxies with AC$< 6$.  Galaxies with classes in the range 6-12 are all concentrated in the high luminosity range. These plots indicate that (a) grand-design spirals tend to have shallow radial abundance gradients (whether they have  a bar or not), (b) low-luminosity unbarred galaxies are all flocculent or multi-arm (AC$<6$), and show a tendency to have steep slopes, especially for the N/O radial profile. Further data in the low-luminosity range would be desirable to draw further conclusions.

%\col{The 'all' you use here can possibly be criticized. In Fig. 12 (right) there are 2 unbarred galaxies that are both the faintest unbarred galaxies and have the steepest O/H gradient. But between MB = -18.5 and -19 there are two more unbarred galaxies, one has a flat gradient, the other an 'intermediate' one. However, what you say seems to apply to the N/O gradient slope. The low number of galaxies available at low luminosity is a limiting factor in the conclusions one can draw. The situation resembles Fig 11 of my 2019 paper, which you introduced earlier.}

Our results then support the hypothesis that gas flows induced by grand-design spirals can also contribute to flatten radial abundance profiles. However, the fact that in the high luminosity range barred and unbarred galaxies with a variety of arm classes coexist (AC 6 up to 12), and all have shallow radial abundance profiles, points to the need for additional flattening agents \citep[e.g. interactions][]{RupkeKewleyChien2010}. These, together with evolutionary effects (Sect.~\ref{alpha_OH_MB}) and contributions from bar- and/or  spiral-induced gas flows could help explaining the observed shallow profiles of the high mass range, and solve the discrepancy  between predicted and observed slopes in unbarred galaxies \citep[e.g.][]{Bresolin2019,Kaplan2016,Pilkington2012}. Additional theoretical and observational work is needed to derive further conclusions.

%%%%%%%%%%%%%%%%%%%%%%%%%%%%%%%%%%%%%%%%%%%%%%%%%%%%%%%%%%%%%%%%%%%%%%%%%%%%%%%%%%%%%%%%%%%%%%%%%%%%%%%%%%%%%%
%%%%%%%%%%%%%%%%%%%%%%%%%%%%%%%%%%%%%%%%%%%%%%%%%%%%%%%%%%%%%%%%%%%%%%%%%%%%%%%%%%%%%%%%%%%%%%%%%%%%%%%%%%%%%%
%%%%%%%%%%%%%%%%%%%%%%%%%%%%%%%%%%%%%%%%%%%%%%%%%%%%%%%%%%%%%%%%%%%%%%%%%%%%%%%%%%%%%%%%%%%%%%%%%%%%%%%%%%%%%%
\section{Summary and conclusions}
\label{conclusiones}
%{\em This section still needs some work.}

This paper is the second one, of a series of two, devoted to revisit the issue of whether galactic bars affect the radial distribution of metals in the gas-phase of  spirals. The project is aimed at solving the existing discrepancies between papers published in the 1990s (with resolved spectroscopy but reduced galaxy samples) and more recent, mostly IFU-based, work (with lower spatial resolution but larger data samples). In \citetalias{paperI} we presented our compiled sample of  \hii\ regions that belong to a sample of 51 nearby spiral galaxies. The sample contains 2831 emission-line fluxes. The chemical abundance ratios of O/H and N/O were derived for all \hii\ regions with a homogeneous methodology that used a variety of strong-line methods (based on R$_{23}$, N2O2, O3N2, N2, HCM, and the $R$ calibration for $12+\log$(O/H), and for $\log$(N/O) on N2O2, N2S2, the $R$ calibration  and HCM). The structural parameters of bars and discs and the radial abundance profiles were also derived with the same methodology for all galaxies. In this paper we analyse the radial $12+\log$(O/H) and $\log$(N/O)  abundance profiles of barred and unbarred galaxies separately. Our main results and conclusions are summarised below:

\begin{itemize}

\item We find no significant difference in central  $12+\log$(O/H) and $\log$(N/O) abundances (as derived from the y-intercepts of the radial profile linear fitting) for strongly barred and unbarred galaxies. However, the averaged $\log$(N/O) abundance of the inner  region (r$\lesssim$2~kpc)  is larger in barred than in unbarred galaxies, in agreement with \citet{bulbos}. % (Figs.~\ref{NO_vs_M},  \ref{empi_vs_M}  and \ref{inner_abund}).

\item The distributions of radial abundance slopes are statistically different for barred and unbarred galaxies for both $12+\log$(O/H) and $\log$(N/O). Barred galaxies have shallower gradients than unbarred galaxies, with the differences being more evident when slopes are expressed in dex~kpc\me\ and dex~$r_e$\me. A deeper look at the data shows that the second parameter is luminosity:  %(Figs.~\ref{NO_vs_M} and \ref{empi_vs_M}):
\begin{itemize}
\item High luminosity strongly barred galaxies show rather shallow $12+\log$(O/H) and $\log$(N/O) radial abundance gradients (in dex~kpc$^{-1}$) with any diagnostic. Although with low number statistics, a shallow radial gradient is also measured for lower luminosity strongly barred galaxies. Therefore, our results suggest that  strongly barred galaxies of any luminosity ($-17\gtrsim$M$_B\gtrsim-22$) have shallow $12+\log$(O/H)  and $\log$(N/O) radial abundance gradients (in dex~kpc$^{-1}$).

\item Unbarred galaxies are indistinguishable in terms of abundance slope from strongly barred galaxies, if we restrict the sample to the brightest (more massive) galaxies (M$_B\lesssim-19.5$). In less luminous (less massive) unbarred spirals the radial abundance profiles in $\log$(N/O) and $12+\log$(O/H) become steeper  with decreasing luminosity  (or mass). 
\end{itemize}

\item The results listed above do not depend either on the presence of {\em breaks} in the radial abundance profiles of the galaxies, or on the exclusion of the \hii\ regions located within stellar bars of the barred galaxies from the radial profile fits.  The fact that the inclusion/exclusion of \hii\ regions located within bars does not affect the results implies that if the observed flattening is produced by the bar, its effect might be a large scale phenomenon that homogenises the gas chemical composition in the whole galactic disc, and  is not limited to the bar area.

\item No correlation is found between radial abundance gradients and current bar properties (length or ellipticity). Therefore, if the radial abundance profile flattening observed in barred galaxies is due to the mixing produced by the bars, it seems that any bar, provided that it has considerable ellipticity  ($e_{bar,d}\gtrsim0.5$) or length ($r_{bar,d}/r_e \gtrsim 0.6$), is able to flatten the abundance profile, with no dependence of the observed gradient on the current bar properties.

\item From the comparison with chemical evolution models and simulations, and with observational work from different authors we conclude:

\begin{itemize}

\item The galaxy luminosity range sampled by different investigations might be the key element for understanding discrepancies in the results, but we can not discard contributions to this discrepancy due to the use of different methodologies for bar classification and/or abundance diagnostics.
  
%Given the luminosity-dependence of the barred/unbarred differences in abundance gradients 
\item Our measured 12$+\log$(O/H) and $\log$(N/O) radial gradients are in agreement with the findings from previous authors and from the predictions of models for a
dependence of the gradient (in units of dex~kpc$^{-1}$) on  the galaxy luminosity. In view of our results it is tempting to go a step forward, as it seems  that this dependence is dominated by unbarred galaxies. 
%We are  aware that our sample is limited in size and a larger galaxy sample would be necessary   in order to corroborate the results presented here.

\item Our results are also in agreement with simulations of Milky Way-mass barred galaxies, that predict a little effect of the bar on the metallicity profile, as no difference in radial abundance gradient is seen among barred and unbarred galaxies in massive/luminous galaxies ($M_*\gtrsim10^{10.4}$~M$_\odot$ or M$_B\lesssim-19.5$). However, a stronger flattening is predicted when a strong bar forms in a galaxy with a steeper initial abundance gradient \citep{Kubryk2013}, as is the case for the lower-mass/luminosity unbarred spirals.

\item The fact that strongly barred galaxies, for which gas flows are known to be relevant, show a shallow present-day radial abundance in  the whole luminosity/mass range, makes  us  agree with other authors \citep[e.g.][]{Bresolin2019,PortinariChiosi1999} in that radial gas flows might be a relevant ingredient in chemical evolution models, especially  for lower-mass (those with a steeper abundance profile) galaxies.  
\end{itemize}
\end{itemize}

Taken all together, our results point to a more efficient flattening  effect of strong bars in lower mass/luminosity galaxies with respect to high mass galaxies. Given the observational result that high luminosity galaxies have the same (shallow) slopes, it is possible that  evolutionary processes alone, but that include  mixing (e.g. by the grand design spiral arms) can explain the flattening at the high mass end. The bar effect in high mass galaxies may be not noticeable due to the presence of an already flat radial initial abundance profile due to the mentioned evolutionary process, as in the high-mass unbarred spirals. Further observational and theoretical work is necessary to make further progress.

This paper re-opens, from an observational point of view, the issue of the potential effects of stellar bars on  the gas-phase of spirals, that recent work based on IFS-data and large data samples suggested was closed.

%Taken all together, our results can be explained with scenario (i) in Section~\ref{scenarios}: The tight relation among  12$+\log$(O/H) and $\log$(N/O) radial gradients with M$_B$ is a consequence of the evolution (inside-out disc growth). Low-luminosity (mass) barred galaxies do not follow this trend, showing shallower abundance profiles as a result of the bar induced gas flows, that are more efficient flattening for low mass discs.

%%%%%%%%%%%%%%%%%%%%%%%%%%%%%%%%%%%%%%%%%%%%%%%%%%%%%%%%%%%%%%%%%%%%%%%%%%%%%%%%%%%%%%%%%%%%%%%%%%%%%%%%%%%%%%%%%
%%%%%%%%%%%%%%%%%%%%%%%%%%%%%%%%%%%%%%%%%%%%%%%%%%%%%%%%%%%%%%%%%%%%%%%%%%%%%%%%%%%%%%%%%%%%%%%%%%%%%%%%%%%%%%%%%

\section*{Acknowledgements}
We acknowledge Sim\'on D\' iaz-Garc\' ia for useful discussions.  We thank Calar Alto Observatory for allocation of director's discretionary time to this programme. We also thank the anonymous referee for his/her suggestions that improved the clarity of the manuscript. AZ, EF and IP acknowledge support from the Spanish {\em Ministerio de Economia y Competitividad} and {\em FEDER programme} via grant AYA2017-84897-P and from the Junta de Andalucia local government through the FQM-108 project. EPM acknowledges  funding from the Spanish MINECO project  Estallidos 6 AYA2016-79724-C4 and  the Spanish Science Ministry "Centro de Excelencia Severo Ochoa Program under grant SEV-2017-0709''.

%%%%%%%%%%%%%%%%%%%%%%%%%%%%%%%%%%%%%%%%%%%%%%%%%%
%%%%%%%%%%%%%%%%%%%%%%%%%%%%%%%%%%%%%%%%%%%%%%%%%%%%%%%%%%%%%%%%%%%%%%%%%%%%%%%%%%%%%%%%%%%%%%%%%%%%%%%%%%%%%%%%%
\section*{Data Availability}
The data underlying this article are available in the article, in \citet{paperI}  and in the sources specified there.

%%%%%%%%%%%%%%%%%%%% REFERENCES %%%%%%%%%%%%%%%%%%

% The best way to enter references is to use BibTeX:

\bibliographystyle{mnras}
\bibliography{referencias} % if your bibtex file is called example.bib

%%%%%%%%%%%%%%%%%%%%%%%%%%%%%%%%%%%%%%%%%%%%%%%%%%
%%%%%%%%%%%%%%%%%%%%%%%%%%%%%%%%%%%%%%%%%%%%%%%%%%
%%%%%%%%%%%%%%%%%%%%%%%%%%%%%%%%%%%%%%%%%%%%%%%%%%
%%%%%%%%%%%%%%%%%%%%%%%%%%%%%%%%%%%%%%%%%%%%%%%%%%
%%%%%%%%%%%%%%%%% APPENDICES %%%%%%%%%%%%%%%%%%%%%
% Don't change these lines
%\bsp	% typesetting comment
%\label{lastpage}
\end{document}